\documentclass[ejs,preprint]{imsart}

\usepackage{url,color}
\usepackage{graphicx}
\usepackage{amsmath, amsthm, amssymb}
\usepackage{algpseudocode}
\usepackage{algorithm}
\usepackage{subfigure}
\usepackage{comment}
\RequirePackage[round]{natbib}
\RequirePackage[colorlinks,citecolor=blue,urlcolor=blue]{hyperref}

\usepackage{multirow}

\newtheorem{theorem}{Theorem}
\newtheorem{lemma}[theorem]{Lemma}

\newtheorem{corollary}[theorem]{Corollary}

\let\hat\widehat

\theoremstyle{remark}
\newtheorem{remark}{Remark}

\newcommand\R{\mathbb{R}}
\newcommand\K{\mathbb{K}}
\newcommand\E{\mathbb{E}}

\newcommand\cL{{\cal L}}

\newcommand\cE{{\cal E}}

\catcode`@=11
\newskip\beforeproofvskip
\newskip\afterproofvskip
\beforeproofvskip=\medskipamount
\afterproofvskip=\bigskipamount

\def\prooftag{Proof}
\def\proofskip{\enspace}

\def\proof{\@ifnextchar[{\@@proof}{\@proof}}  
\def\@startproof{\par\vskip\beforeproofvskip\leavevmode}
\def\@proof{\@startproof{\scshape\prooftag.}\proofskip}
\def\@@proof[#1]{\@startproof {\scshape\prooftag #1.}\proofskip}

\catcode`@=12

\let\hat\widehat
\let\tilde\widetilde

\newenvironment{enum}{
\begin{enumerate}
  \setlength{\itemsep}{1pt}
  \setlength{\parskip}{0pt}
  \setlength{\parsep}{0pt}
}{\end{enumerate}}

\begin{document}

\begin{frontmatter}

\title{{\color{black}Importance Sampling and its Optimality} for Stochastic Simulation Models}
\runtitle{Stochastic Importance Sampling}

\begin{aug}
  \author{Yen-Chi Chen\ead[label=e1]{yenchic@uw.edu}}
  \and
  \author{Youngjun Choe\ead[label=e2]{ychoe@uw.edu}}
  
    \address{University of Washington,\\
    Department of Statistics\\
  Seattle, WA 98195 \\
           \printead{e1}\thanksref{T1}}

    \address{University of Washington,\\
    Department of Industrial and Systems Engineering\\
  Seattle, WA 98195 \\
           \printead{e2}\thanksref{T1}}

  \thankstext{T1}{Both authors equally contributed to this work.}

  \runauthor{Chen and Choe}

\end{aug}

\begin{abstract}
We consider the problem of estimating an expected outcome from a stochastic simulation model.
Our goal is to develop a theoretical framework on importance sampling for such estimation. 
By investigating the variance of an importance sampling estimator,
we propose a two-stage procedure that involves a regression stage and a sampling stage
to construct the final estimator.
We introduce a parametric and a nonparametric regression estimator in the first stage and study how the allocation
between the two stages affects the performance of the final estimator. 
We analyze the variance reduction rates
and derive
oracle properties of both methods. 
We evaluate the empirical performances of the methods using two numerical examples and a case study on 
wind turbine reliability evaluation.
\end{abstract}




\end{frontmatter}

\section{Introduction}\label{sec::intro}

The 2011 Fisher lecture \citep{wu2015post} features the landscape change in engineering, where computer simulation experiments are replacing physical experiments thanks to the advance of modeling and computing technologies. An insight from the lecture highlights that traditional principles for physical experiments do not necessarily apply to virtual experiments on a computer. The virtual environment calls for new modeling and analysis frameworks distinguished from those developed under the constraint of physical environment. In this context, this study considers a new problem that emerged along with the recent development of stochastic simulation-based engineering.

A concrete motivating problem of this study is estimating a system failure probability based on stochastic simulations (although our methods are more generally applicable to the estimation of any expected outcome, as detailed in the next section). A system configuration, $X$, is randomly sampled from a known density $p$ and passed on to a stochastic simulation model. 
The simulation model, regarded as a stochastic black-box, produces $V$ that follows an unknown distribution depending on $X$. 
When $V$ is greater than a threshold, say $\xi$, the system fails. 
Thus, the goal is to estimate the probability $P(V\geq \xi)$ when $X$ is from the density $p$.
Specifically, our case study uses the simulation model that is designed to mimic the real system accurately. Thus, it takes roughly 1-min wall-clock time to simulate 10-min real operation of the system on a computer commonly available nowadays. The engineering goal is estimating the probability of system failure during 50-year operation, which is computationally challenging even with U.S. national labs' supercomputers \citep{Manuel2013,graf2016high}. 

Such computational challenges are commonly observed in engineering simulations. Finite element simulations, which are used widely in various engineering applications, can take hours of computing time to obtain a single data point \citep[e.g.,][]{qian2006building}. Despite the computational expense, highly accurate simulations are cost-effective alternatives to physical experiments and used widely in industry (e.g., Ford Motor Company's crash simulation \citep{wang2007review}) and in government (e.g., NASA's rocket booster simulation \citep{gramacy2012bayesian}).

The overarching goal of this study is to propose a framework on analyzing the problem of minimizing the necessary computational burden while maintaining the same level of estimation accuracy. A flip-side of the same problem is minimizing the estimation variance given fixed computational resource.  Variance reduction techniques (VRTs) in the simulation literature aim to reduce the variance of estimator in simulation experiments. Traditional VRTs are well studied for the simulation model that outputs $V$ given $X$ in a deterministic fashion, also known as the deterministic simulation model (see Chapter~9 of \citet{kroese:2011handbook} for survey of such VRTs), when the input $X$ is sampled from a known distribution. For stochastic simulation models, if their underlying processes have known properties (e.g., Markovian), \citet{glynn1989importance} and \citet{heidelberger1995fast} provide VRTs.  For black-box stochastic simulation models, few studies \citep[e.g.,][]{choe2015importance,graf2017advances} consider VRTs. The research on VRTs for block-box stochastic simulations is still underdeveloped 
despite the rapid growth of such simulations being used in real systems, for example, chemical systems \citep{gillespie2001approximate}, biological systems \citep{henderson2012bayesian}, and engineering systems \citep{ankenman2010stochastic,picheny2013quantile,plumlee2014building}.  

Among VRTs, importance sampling \citep{kahn1953} is known to be one of the most effective methods and has been used widely in various applications such as communication systems \citep{heidelberger1995fast,bucklew2004}, finance \citep{owen2000safe,glasserman2005importance}, insurance \citep{blanchet2011importance}, and reliability engineering \citep{au1999new,lawrence2013model,choe2016} to name a few.

In the vast majority of literature, importance sampling takes a parametric form tailored to a problem at hand for both deterministic simulation model \citep[e.g.][]{lawrence2013model} and stochastic counterpart \citep{choe2015importance}. 
Nonparametric approaches are also proposed for deterministic simulation models \citep{zhang1996nonparametric,neddermeyer2009computationally}. To the best of our knowledge, no nonparametric approach is developed for stochastic simulation models. This study particularly considers the black-box stochastic simulation model whose output takes an unknown stochastic relationship with the input. 


In this paper, we focus on the theoretical aspects of importance sampling with stochastic simulation models. 
We design two-stage sampling methods that may perform as good
as the best sampling method (also known as the oracle property). 
The main contributions of this paper to the existing body of literature are as follows:
\begin{itemize}
	\item We introduce an importance sampling approach to estimate the expectation of black-box stochastic simulation output and study the optimal importance sampler (Section~\ref{sec::is_stoch}).
	\item We design a two-stage procedure that uses a parametric or a nonparametric regression estimator to approximate the optimal importance sampler (Figure~\ref{fig::parametric} and \ref{fig::nonparametric}).
	\item We analyze the allocation of the resources in both stages (Theorem~\ref{thm::P_var} and \ref{thm::NP_var}) and study the convergence of the two-stage procedure toward the oracle importance sampler (Corollary~\ref{cor::Pmn} and \ref{cor::mn}).
	\item We conduct an extensive numerical study to evaluate empirical performances of the two-stage importance samplers (Section~\ref{sec::ex1} and \ref{sec::ex2}).
	\item We apply our methods to a case study on wind turbine reliability evaluation (Section~\ref{sec::emp::case_study}) to validate our results.
\end{itemize}

This paper is organized as follows.
In Section~\ref{sec::is_stoch} we formulate the stochastic simulation-based estimation problem and introduce a two-stage procedure to
estimate the expected simluation output. 
In Section~\ref{sec::thm} we study the theoretical performance of the proposed procedure
and derive the corresponding oracle properties. 
In Section~\ref{sec::emp} we evaluate the empirical performance of our approach using two numerical examples 
and a wind turbine simulator. 
We discuss our result in Section~\ref{sec::disc}.

\section{Importance Sampling for the Stochastic Simulation Model}\label{sec::is_stoch}
A stochastic simulation model 
takes an input configuration value and then returns a random number representing the outcome
of this simulation result.
The input configuration determines the distribution of the (random) outcome. 
Thus, the outcome of a stochastic simulation model can be represented
by a random variable $V$ conditioned on the input configuration $x$
and the CDF of $V$ is
$$
F_{V|{\sf config} = x}(v|x) = P(V\leq v|{\sf config} = x),
$$
where ${\sf config} = x$ denotes choosing the configuration to be $x$. 
For simplicity, we denotes the random variable
$V$ conditioned on ${\sf config} = x$ as $V(x)$.



In many scientific or engineering problems \citep[e.g.,][]{heidelberger1995fast,au2003, bucklew2004,graf2017advances}, 
we assume the nature generates the configuration from a known density $p$ and
we are interested in evaluating the quantity 
\begin{equation}
\cE = \mathbb{E}(g(V(X))) = \int \mathbb{E}(g(V)|{\sf config}=x) p(x)dx,
\label{eq::target}
\end{equation}
where $g $ is a known function.

{\bf Example 1.} A common example for equation \eqref{eq::target} is the case when $g(v) = v$, often considered in the literature on two-level nested simulation \citep{sun2011efficient}, where the outer level generates a scenario (or configuration) according to a known density $p$ and conditioning on the scenario, the inner level simulates a random outcome whose mean is of interest. Applications include decision theory \citep{brennan2007calculating}, financial engineering \citep{Staum2009}, and queuing system \citep{sun2011efficient}.


{\bf Example 2.}
Another example takes $g(v) = 1(v \in S_\xi)$ for some set $S_\xi$ parametrized by $\xi$.
Specifically,  \citet{choe2015importance} considers a reliability evaluation problem where $V$ stands for an instability measurement of a system that fails when $V$ falls into $S_\xi = \left[\xi, \infty\right)$.
The goal is to estimate the failure probability when the system is exposed to the nature. 
In the natural environment, the configuration behaves like a random variable from a density $p$.

%


To estimate $\cE$, we can choose several configurations $x_1,\cdots, x_n$
and then run the simulation to obtain realizations $v_1=V(x_1), \cdots, v_n = V(x_n)$. 
However, 
generating $V$ from a given configuration $x$ is often computationally {\color{black}expensive}
for stochastic simulation models. 
Therefore, we would like to run the simulation as few as possible.
To put this constraint into consideration,
we assume that
we run the simulation only $n$ times but we are able to choose the configuration
for each simulation. 
We choose $n$ configurations and evaluate the corresponding value of $V$.
Namely, we only have pairs $(x_1,V_1),\cdots, (x_n,V_n)$, where 
each $V_i$ is a realization of the random variable $V(x_i)$.
Such a constraint on the number of simulations, $n$, is sometimes called a \emph{computational budget}. 

%


Under such situation, a natural question is: \emph{How do we choose the configurations $x_1,\cdots,x_n$?}
Here we use the idea from importance sampling --
we choose $x_1,\cdots,x_n$ from a density function $q$.
In other words, we first sample $X_1,\cdots,X_n$ from $q$
and then use $x_i=X_i$ as the configuration to run the $i$-th stochastic simulation. 
The density $q$ is called \emph{sampling density}.
Note that each configuration does not necessarily have to be from the same density function.

When we generate $X_1,\cdots,X_n$ from $q$
and then obtain $V_1,\cdots, V_n$ accordingly, a simple estimator of $\cE$ is
\begin{equation}
\hat{\cE}_q = \frac{1}{n}\sum_{i=1}^n g(V_i) \frac{p(X_i)}{q(X_i)}.
\label{eq::est1}
\end{equation}
It is easy to see that $\hat{\cE}_q$ is an unbiased estimator under the assumption that $q(x)=0$ implies $g(V(x))p(x)=0$ for all $x$,
i.e.,
$$
\mathbb{E}\left(\hat{\cE}_q\right) = \cE
$$
when the support of $q$ covers the support of $g(V(x))p(x)$.
We call this type of estimator an \emph{importance sampling estimator}. 
Throughout this paper, we will focus on importance sampling estimators.


Using an importance sampling estimator \eqref{eq::est1},
a key problem we want to address is:
\emph{what will be the optimal sampling density $q^*$ that minimizes the estimation error? }
Because the estimator \eqref{eq::est1} is unbiased, we only need to find the minimal variance estimator.
Thus, the above question is equivalent to:
\emph{what will be the optimal sampling density $q^*$ that minimizes the variance of $\hat{\cE}_q$?}
The following lemma provides an answer to the above questions:
\begin{lemma}
Assume $X_1,\cdots,X_n$ are IID from a density function $q$.
Let $r^\dagger(x) = \E(g(V(X))|X=x)$ and $r(x) = \E(g^2(V(X))|X=x)$. 
Then variance of $\hat{\cE}_q$ equals to 
\begin{align}
{\sf Var} \left(\hat{\cE}_q\right) &=  \frac{1}{n} \left(\mathbb{E}_{X_i\sim q}\left(r(X_i) \frac{p^2(X_i)}{q^2(X_i)}\right) - \mathbb{E}^2(r^\dagger(X^*))\right) \nonumber\\
&\geq \frac{1}{n} \left(\mathbb{E}^2\left(\sqrt{r(X^*)}\right)-\mathbb{E}^2(r^\dagger(X^*))\right) \nonumber\\
& \equiv \frac{1}{n} V_{\min}  \label{eq:V_min}
\end{align}
where $X^*$ is a random variable from the density $p$.
The equality holds when we choose $q$ to be $q(x) = q^*(x) \propto \sqrt{\E(g^2(V(X))|X=x)} \cdot p(x)$.
Namely, the optimal sampling density is $q^*(x)$.
\label{lem::var}
\end{lemma}

We call the quantity $V_{\min}$ the \emph{oracle variance} of the importance sampling. 
It is the minimal variance that an importance sampler can achieve. 
A widely studied special case in engineering is the deterministic simulation model where ${\sf Var} (V|X=x)=0$ for all $x$, which implies $V_{\min} = 0$ for any nonnegative function $g(v)$ \citep[e.g.,][]{kahn1953, au1999new,kroese:2011handbook}. 
The density that leads to the oracle variance, $q^*$, is called the optimal sampling density. This density is a modification from the natural configuration density $p$;
$q^*$ puts more weight on the regions with higher $r(x)$ (e.g., higher probability of system failure).
{\color{black}As long as $r$ and $r^\dagger$ are uniformly bounded within the support of $p$, the variance is finite. }

However, we cannot directly generate configurations from $q^*$
because it involves the unknown quantity $r(x)=\E(g^2(V(X))|X=x)$.
A remedy to this problem is to apply a \emph{two-stage sampling}. 
In the first stage, we generate part of configurations and evaluate the corresponding values of 
$V(x)$.
Using the sample in the first stage, we obtain a pilot estimator $\hat{r}$ of $r$.
In the second stage, we generate configurations based on an estimate of $q^*$ using the pilot estimator $\hat{r}$
and use the remaining computational budget to evaluate values of $V(x)$.
Finally, we use samples from both stages to form the final estimator of $\cE$.


Here is a useful insight in estimating $r(x)$.
Let $Y$ be a random variable such that $Y = g^2(V(X))$.
Then $r(x) = \E(g^2(V(X))|X=x)=\E(Y|X=x) $.
Thus, estimating $r(x)$ is equivalent to estimating the regression function with 
the observations $(X_1,Y_1 = g^2(V_1)), \cdots, (X_n,Y_n = g^2(V_n))$ assuming we have $n$ observations.

Thus the two-stage procedure can be summarized as follows.
We first generate a size $m$ sample 
$$
(X_1,V_1),\cdots,(X_m,V_m)
$$
where $X_i, i = 1,\cdots,m$, are from an initial sampling density $q_0$.
Then we transform $V_i$ into $Y_i = g^2(V_i)$, which leads to a sample
$$
(X_1,Y_1),\cdots,(X_m,Y_m).
$$
Now we estimate the regression function $r(x)$ by a regression estimator $\hat{r}(x)$ and compute the corresponding estimator $\hat{q}^*$
of the oracle sampling density $q^*$.
Finally, we generate the remaining data points 
$$
(X_{m+1},V_{m+1}),\cdots,(X_n,V_n)
$$
from $\hat{q}^*$ and pool both samples together to form the final estimator of the quantity $\cE$.
Because $\hat{q}^*$ will tend to be closer to $q^*$ compared to the initial sampling density $q_0$,
estimating $\cE$ using the sample in the second stage is more efficient (lower variance). 

The sample size $m$ in the first stage is a crucial quantity in our analysis.
The quantity $m$ is called the \emph{allocation}.
When $m$ is too small, the estimator of $q^*$ is inaccurate,
so that the overall estimation efficiency is suboptimal. 
When $m$ is too large, we only have a small budget for 
the second stage so that the overall estimation efficiency is low as well. 
As a result, to balance the estimation accuracy of $q^*$ and the size of efficient sample in the second stage,
there will be an optimal value of $m$ depending on the total sample size $n$.
In what follows we propose two different models to estimate $r$ and $q^*$
and analyze the optimal value of the allocation $m$.

\subsection{Parametric Importance Sampling}

As in the regression analysis, a straightforward approach of estimating the regression
function is to assume a parametric model and estimate the corresponding parameters.
Namely, we assume $r(x) = r_\theta(x)$ for some $\theta\in\Theta$ 
and use the first part of the sample to estimate $\theta$.

To estimate $r_\theta(x)$, we use a classical approach--\emph{the least square method}:
\begin{equation}
\hat{\theta}_m = {\sf argmin}_{\theta \in \Theta} \sum_{i=1}^m \|Y_i-r_\theta(X_i)\|^2.
\label{eq::LS}
\end{equation} 
Then the estimator $\hat{r}(x) = r_{\hat{\theta}_m}(x)$.
Note that one can also assume a parametric form for the distribution of $Y_i|X_i$ and 
then use a maximum likelihood estimator. 
Using the estimator $\hat{\theta}_m$, we can then estimate the regression function $r_{\hat{\theta}_m}$
and
construct the estimated optimal sampling density 
$$
q^*_{\hat{\theta}_m}(x) \propto\sqrt{r_{\hat{\theta}_m}(x)}\cdot p(x).
$$
For the remaining $(n-m)$ data points, we generate the configurations from $q^*_{\hat{\theta}_m}$,
run the simulation,
and estimate $\cE$ accordingly. 


\begin{figure}[h]
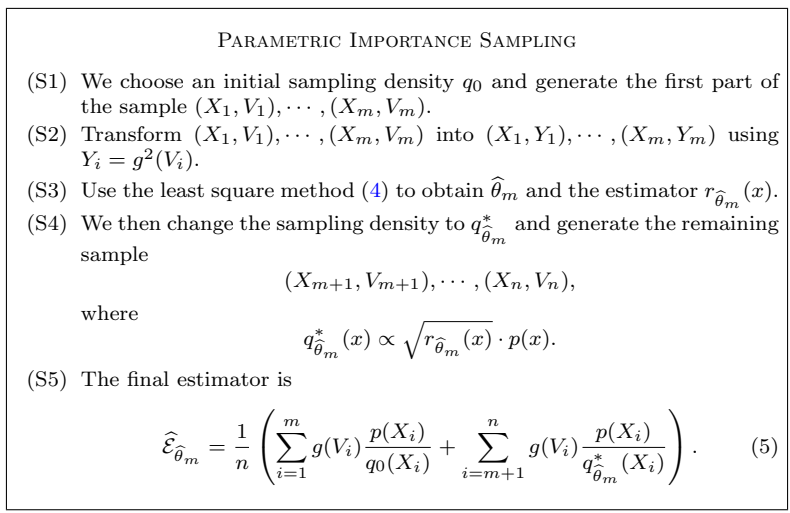

\fbox{\parbox{4in}{
\begin{center}
{\sc Parametric Importance Sampling}
\end{center}
\begin{center}
\begin{enum}
\item[(S1)] We choose an initial sampling density $q_0$ and generate the first part of the sample $(X_1,V_1),\cdots, (X_m,V_m)$.
\item[(S2)] Transform $(X_1,V_1),\cdots, (X_m,V_m)$ into $(X_1,Y_1),\cdots, (X_m,Y_m)$ using $Y_i = g^2(V_i)$.
\item[(S3)] Use the least square method \eqref{eq::LS} to obtain $\hat{\theta}_m$ and the estimator $r_{\hat{\theta}_m}(x)$.
\item[(S4)] We then change the sampling density to $q^*_{\hat{\theta}_m}$ and generate the remaining sample 
$$(X_{m+1},V_{m+1}),\cdots, (X_n,V_n),$$ 
where
$$
q^*_{\hat{\theta}_m}(x) \propto\sqrt{r_{\hat{\theta}_m}(x)}\cdot p(x).
$$
\item[(S5)] The final estimator is 
\begin{equation}
\hat{\cE}_{\hat{\theta}_m} = \frac{1}{n}\left(\sum_{i=1}^m g(V_i)\frac{p(X_i)}{q_0(X_i)} + 
\sum_{i=m+1}^n g(V_i)\frac{p(X_i)}{q^*_{\hat{\theta}_m}(X_i)}\right). \label{eq:param_IS_estimator}
\end{equation}
\end{enum}
\end{center}
}}
\caption{Parametric importance sampling for the stochastic simulation model.}
\label{fig::parametric}
\end{figure}

We summarize our Parametric Importance Sampling method in Figure~\ref{fig::parametric}.
Later in Section~\ref{sec::thm::parametric} we will derive the variance of
this approach and show that the optimal allocation is to choose $m = O(n^{\frac{2}{3}})$.



\subsection{Nonparametric Importance Sampling}	\label{sec::NIS}


Now we consider estimating $r(x)$ nonparametrically.
For simplicity, we use the kernel regression \citep{nadaraya1964estimating,watson1964smooth}.
Note that other nonparametric regression approach, such as the local polynomial regression \citep{wasserman2006all}, also works.
The kernel regression uses the estimator
\begin{equation}
\begin{aligned}
\hat{r}_h(x) = \frac{\sum_{i=1}^m Y_i K\left(\frac{x-X_i}{h}\right)}{\sum_{i=1}^m K\left(\frac{x-X_i}{h}\right)},
\end{aligned}
\label{eq::est2}
\end{equation}
where $K$ is a smooth function (known as the kernel function) such as a Gaussian, and $h>0$ is the smoothing bandwidth.
Similar to the parametric approach, 
we then use this estimator to construct an estimated optimal sampling density 
$$
\hat{q}^*_h(x) \propto \sqrt{\hat{r}_h(x)} \cdot p(x),
$$
generate the remaining data points from it,
and construct the final estimator using the procedure described previously.


\begin{figure}[h]
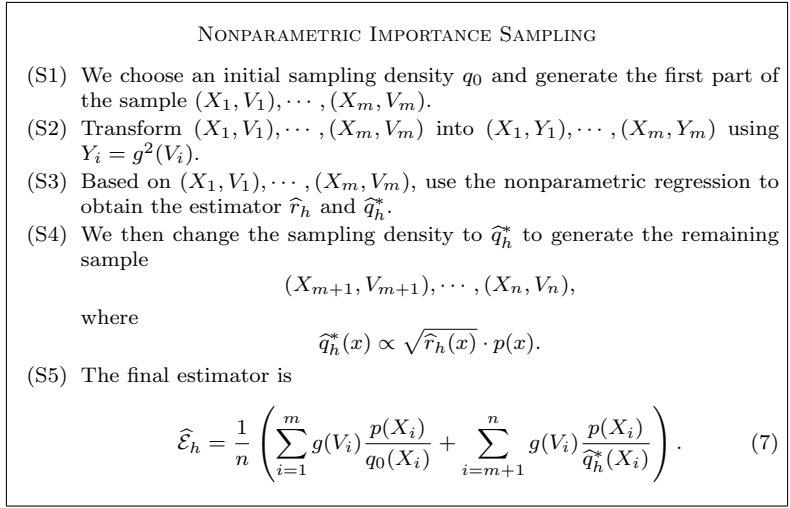

\fbox{\parbox{4in}{
\begin{center}
{\sc Nonparametric Importance Sampling}
\end{center}
\begin{center}
\begin{enum}
\item[(S1)] We choose an initial sampling density $q_0$ and generate the first part of the sample $(X_1,V_1),\cdots, (X_m,V_m)$.
\item[(S2)] Transform $(X_1,V_1),\cdots, (X_m,V_m)$ into $(X_1,Y_1),\cdots, (X_m,Y_m)$ using $Y_i = g^2(V_i)$.
\item[(S3)] Based on $(X_1,V_1),\cdots, (X_m,V_m)$, use the nonparametric regression to obtain the estimator $\hat{r}_{h}$ and $\hat{q}^*_h$. 
\item[(S4)] We then change the sampling density to $\hat{q}^*_h$ to generate the remaining sample 
$$(X_{m+1},V_{m+1}),\cdots, (X_n,V_n),$$ 
where
$$
\hat{q}^*_h(x) \propto \sqrt{\hat{r}_h(x)} \cdot p(x).
$$
\item[(S5)] The final estimator is 
\begin{equation}
\hat{\cE}_h = \frac{1}{n}\left(\sum_{i=1}^m g(V_i)\frac{p(X_i)}{q_0(X_i)} + 
\sum_{i=m+1}^n g(V_i)\frac{p(X_i)}{\hat{q}^*_h(X_i)}\right). \label{eq:nonparam_IS_estimator}
\end{equation}
\end{enum}
\end{center}
}}
\caption{Nonparametric importance sampling for the stochastic simulation model.}
\label{fig::nonparametric}
\end{figure}

Figure~\ref{fig::nonparametric} summarizes the procedure of nonparametric importance sampling. 
There are two tuning parameters we need to select: the smoothing bandwidth $h$ and the allocation size $m$.
The smoothing bandwidth can be chosen by either cross-validation or a reference rule.
In Section~\ref{sec::thm::nonparametric}, we will derive the optimal rate for the smoothing bandwidth 
 $h = O\left(\left(\frac{\log m}{m}\right)^{\frac{1}{d+4}}\right)$
and the optimal allocation rate $m = O\left(\left(\frac{n}{\log n}\right)^{\frac{d+4}{d+6}}\right)$.


{\bf Remark.}
We may rewrite  $\hat{\cE}_h = \hat{\cE}_1 + \hat{\cE}_2$,
where $\hat{\cE}_1 = \frac{\sum_{i=1}^m g(V_i)\frac{p(X_i)}{q_0(X_i)}}{n}$
and $\hat{\cE}_2 = \frac{\sum_{i=m+1}^n g(V_i)\frac{p(X_i)}{\hat{q}^*_h(X_i)}}{\color{black}n}$
are the estimator constructed from the two samples. 
If we know the variance of $\hat{\cE}_1$ and $\hat{\cE}_2$,
then we can further reduce the variance of the final estimator by
doing a weighted combination $\hat{\cE}_h^\alpha = \alpha\hat{\cE}_1 + (1-\alpha)\hat{\cE}_2$,
with $\alpha = \frac{{\sf Var}(\hat{\cE}_2)}{{\sf Var}(\hat{\cE}_1)+{\sf Var}(\hat{\cE}_2)}$.
One can easily show that ${\sf Var}(\hat{\cE}_h^\alpha)< {\sf Var}(\hat{\cE}_h)$ an $\E(\hat{\cE}_h^\alpha) = \cE$
so $\hat{\cE}_h^\alpha$ is a better estimator. 


\section{Theoretical Analysis}\label{sec::thm} 
Throughout our analysis, we assume that the natural configuration density $p$ has a compact support $\K\subset \R^d$
and the support of the initial sampling density $q_0$ contains $\K$.

\subsection{Parametric Importance Sampling}	\label{sec::thm::parametric}
{\bf Assumptions.}
\begin{itemize}
\item[(P1)] There exists an unique ${\theta_0}\in\Theta$ such that $r(x) = r_{\theta_0}(x)$ and $\sup_{x\in\K}{\sf Var}(Y_1-r_{\theta_0}(X_1)|X_1=x) \leq \sigma^2_{\max}<\infty$. The support of $r_\theta(x)$ contains the support of $p(x)$ for every $\theta\in\Theta$
and $r(x)>0$ for all $x\in\K$. {\color{black}Also, the PDF $q_0(x)$ exists and has non-zero variance.}
\item[(P2)] Let $\ell(\theta) = \E\|Y_1-r_\theta(X_1)\|^2$. The Hessian matrix $H(\theta)=  \nabla_\theta \nabla_\theta \ell(\theta)$ 
is positive definite at $\theta \in B(\theta_0,R_0)$ for some $R_0<\infty$ and $\theta_0$ is the one in (P1).
Note that $B(x,r)$ is a ball centered at $x$ with radius $r$.
\item[(P3)] There exists a positive $L_0<\infty$ such that for any $\theta_1,\theta_2 \in B(\theta_0, R_0)$, 
$$
\sup_{x\in\K}|r_{\theta_1}(x)-r_{\theta_2}(x)|\leq L_0\cdot\|\theta_1-\theta_2\|,
$$
where $\theta_0,R_0$ are defined in (P2).
\end{itemize}

(P1) means that the model is correctly specified--the regression function can be parametrized in the parametric model we consider. 
(P2) is a common assumption in the M-estimation theory \citep{van1996weak,van2000asymptotic} to derive the convergence rate. 
The extra assumption (P3) is a mild assumption that converts the convergence rate of
parameter estimation to the convergence rate of function estimation.
As long as $r_\theta(x)$ is smooth within
an open set around $\theta_0$, (P3) holds.


The following theorem describes the estimation error when the parametric family contains 
the true regression function.
\begin{theorem}
Assume (P1--3).
The error rate for the estimator $r_{\hat{\theta}_m}(x)$ is 
$$
\sup_{x\in\K}\|r_{\hat{\theta}_m}(x)  - r(x)\| = O_P\left(\sqrt{\frac{1}{m}}\right).
$$
\label{thm::Prate}
\end{theorem}
Theorem~\ref{thm::Prate} presents the error rate for estimating $r(x)$ 
when the model is correct. 
Based on this error rate, we can further derive the variance of the parametric importance sampler
in Figure~\ref{fig::parametric}.

\begin{theorem}
Assume (P1--3).
Let $V_{q_0} = \mathbb{E}_{X_i\sim q_0}\left(r(X_i) \frac{p^2(X_i)}{q_0^2(X_i)}\right)-\mathbb{E}^2\left(\sqrt{r(X^*)}\right)$
be the excess variance from using $q_0$ compared to $q^*$. 
The variance of the estimator $\hat{\cE}_{\hat{\theta}_m}$ is
\begin{equation*}
\begin{aligned}
{\sf Var} \left(\hat{\cE}_{\hat{\theta}_m}\right) 
 = \frac{1}{n}V_{\min} + \frac{1}{n^2}\left(m\cdot V_{q_0} + (n-m)\cdot O\left(\sqrt{\frac{1}{m}}\right)\right).
\end{aligned}
\end{equation*}
\label{thm::P_var}
\end{theorem}
Theorem~\ref{thm::P_var} has three components.
The first component $V_{\min}$ is the oracle variance we have mentioned previously.
It is the minimal variance that can be achieved by an importance sampling estimator. 
The second component $\frac{1}{n^2}\cdot m\cdot V_{q_0}$ is the excess variance due
to the initial sampling density. 
The third component $\frac{1}{n^2}\cdot (n-m)\cdot O\left(\sqrt{\frac{1}{m}}\right)$
is the excess variance due to the error of the estimator $r_{\hat{\theta}_m}(x)$.

 
By optimizing $m$ with respect to the second and third components, we obtain
the optimal rate of $m$ as a function of sample size $n$:
\begin{align*}
&m\cdot V_{q_0} = (n-m)\cdot O\left(\sqrt{\frac{1}{m}}\right)\\
\Longrightarrow \quad& m^{\frac{3}{2}} \asymp n\\
\Longrightarrow\quad& m \asymp n^{\frac{2}{3}},
\end{align*}
where the notation $\asymp$ means that the two quantities will be of the same order, i.e., 
$a_n\asymp b_n\Leftrightarrow \lim_{n\rightarrow \infty}\frac{a_n}{b_n} \in (0,\infty)$.
Thus, the optimal allocation is to choose $m  \asymp n^{\frac{2}{3}}$, which leads to the following:
\begin{corollary}
Assume (P1--3).
When  $m  \asymp n^{\frac{2}{3}}$,
the variance of the estimator $\hat{\cE}_{\hat{\theta}_m}$ is
$$
{\sf Var} \left(\hat{\cE}_{\hat{\theta}_m}\right) = \frac{1}{n}V_{\min} \left(1+ O\left( n^{-\frac{1}{3}}\right)\right).
$$
\label{cor::Pmn}
\end{corollary}
That is, if the model is correctly specified, the excess variance 
shrinks at rate $O\left( n^{-\frac{1}{3}}\right)$ under the optimal allocation.

{\color{black}
Note that if we add conditions so that 
$$\frac{q^*_{\hat\theta_m}(x)- q^*(x)}{p(x)}
 = (\hat\theta_m- \theta^*) \cdot C(x)+ o_P(\|\hat\theta_m - \theta^*\|^2)
$$
for some uniformly bounded function $C(x)$,
the variance in Theorem \ref{thm::P_var} can be improved in the sense that the final term will be $(n-m)O\left(\frac{1}{m}\right)$,
which leads to the optimal allocation $m\asymp n^{\frac{1}{2}}$.
This allocation rate is similar
to \cite{li2013two} although we are considering different scenarios. \cite{li2013two} considers mixture importance sampling for a deterministic (as opposed to stochastic) simulation model and study the optimal allocation, where the optimal mixing weight (of each component) is unknown and has to be estimated using a pilot sample.
In our case, the parametric model is to approximate the regression function implied by a stochastic simulation model. 
}

The key assumption of the parametric method is (P1): the actual $r(x)$
belongs to the parametric family.
However, if this assumption is violated, then 
the excess variance in the parametric method will never shrink to $0$.
\begin{theorem}
Assume (P2--3). 
If $r(x) \neq r_\theta(x)$ for all $\theta\in\Theta$, 
the variance of the parametric estimator 
\begin{equation*}
\begin{aligned}
{\sf Var} \left(\hat{\cE}_{\hat{\theta}_m}\right) 
 \geq \frac{1}{n}V_{\min} + \frac{1}{n}V_{\theta^*} 
\end{aligned}
\end{equation*}
where 
\begin{align*}
V_{\theta^*} &= \inf_{\theta\in\Theta}\mathbb{E}\left(r(X_\theta) \frac{p^2(X_\theta)}{q_\theta^2(X_\theta)}\right)-\mathbb{E}^2\left(\sqrt{r(X^*)}\right)>0,\\
X_\theta\sim q_\theta(x)&\propto  \sqrt{r_\theta(x)} \cdot p(x).
\end{align*}
\label{thm::P_inconsistent}
\end{theorem}
The proof of this theorem is trivial, so we omit it.
Theorem~\ref{thm::P_inconsistent} proves that when the model is incorrectly specified, 
there is an additional variance $V_{\theta^*}$ that never disappears.
Thus, the variance of the parametric importance sampler will not converge to 
the optimal variance. 
Later we will see that this implies that the parametric importance sampler does
not have the oracle inequalities when the model is incorrectly specified.

\subsection{Nonparametric Importance Sampling}
In this section, we study the properties of the nonparametric importance sampler in Figure~\ref{fig::nonparametric}.
Similarly as the parametric importance sampler, we first derive the convergence rate of estimating $r(x)$, then derive a variance decomposition for ${\sf Var} \left(\hat{\cE}_h\right)$, and finally 
study the optimal allocation.

{\bf Assumptions.}
\begin{itemize}
\item[(N1)] $\sup_{x\in\K}{\sf Var}(Y_1-r(X_1)|X_1=x) \leq \sigma^2_{\max}<\infty$ and $r(x)>0$ for all $x\in\K$. 
\item[(N2)] For all $x$, the function $r(x)$ has bounded second derivative and $q_0(x)$
has bounded first derivative and $\sup_{x\in\K}q_0(x)\geq q_{\min}>0$.
\item[(K1)] The kernel function $K(x)$ is symmetric and
$$
\int K(x)dx=1,\quad \int \|x\|^2 K(x)dx<\infty,\quad \int   K^2(x)dx<\infty.
$$
\item[(K2)] The collection of functions
\begin{align*}
\mathcal{K} &= \left\{y\mapsto K\left(\frac{x-y}{h}\right): x\in\K, h>0\right\},
\end{align*} 
is a VC-type class. i.e. 
there exists constants $A,v$ and a constant envelope $b_0$ such that
\begin{equation}
\sup_{Q} N(\mathcal{K}, \cL^2(Q), b_0\epsilon)\leq \left(\frac{A}{\epsilon}\right)^v,
\label{eq::VC}
\end{equation}
where $N(T,d_T,\epsilon)$ is the $\epsilon$-covering number for a
semi-metric set $T$ with metric $d_T$ and $\cL^2(Q)$ is the $L_2$ norm
with respect to the probability measure $Q$.
\end{itemize}

(N1) and (N2) are common assumptions for nonparametric regression; see, e.g., 
\cite{wasserman2006all} and \cite{gyorfi2006distribution}.
(K1) is a standard condition on kernel function \citep{wasserman2006all,scott2015multivariate}.
(K2) regularizes the complexity of kernel functions so that 
we have a uniform bound on the stochastic variation.
This assumption
was first proposed in \cite{Gine2002} and \cite{Einmahl2005}
and later was used in various studies such as \cite{genovese2014nonparametric,chen2015asymptotic,chen2017density}.

Based on the above assumptions, 
the uniform convergence rate of the kernel estimator $\hat{r}_h(x)$ is given by the following.
\begin{theorem}
Assume (N1--2), (K1--2). 
The error rate of the kernel estimator $\hat{r}_h(x)$ is 
$$
\sup_{ x\in \K}\|\hat{r}_h(x)  - r(x)\| = O(h^2) +O_P\left(\sqrt{\frac{\log m}{mh^d}}\right).
$$

\label{thm::rate}
\end{theorem}
The error in Theorem~\ref{thm::rate} can be decomposed into two parts:
the bias part $O(h^2)$ and the stochastic variation $O_P\left(\sqrt{\frac{\log m}{mh^d}}\right)$ 
(which is related to the variance).
In many nonparametric studies, 
similar bounds appear for density estimation;
see, e.g., 
\cite{Gine2002,Einmahl2005,genovese2014nonparametric,chen2015asymptotic,chen2016nonparametric}.

By Theorem~\ref{thm::rate},
the optimal bandwidth $h^*  \asymp \left(\frac{\log m}{m}\right)^{\frac{1}{d+4}}$ 
leads to the optimal error rate
\begin{equation}
\hat{r}_{h^*}(x)  - r(x) = O_P\left(\left(\frac{\log m}{m}\right)^{\frac{2}{d+4}}\right).
\label{eq::unif1}
\end{equation}
Under such an optimal error rate, we again obtain the variance decomposition for the nonparametric importance sampler.
\begin{theorem}
Assume (N1--2), (K1--2).
Let $V_{q_0} = \mathbb{E}_{X_i\sim q_0}\left(r(X_i) \frac{p^2(X_i)}{q_0^2(X_i)}\right)-\mathbb{E}^2\left(\sqrt{r(X^*)}\right)$
be the excess variance from using $q_0$ compared to $q^*$. 
The variance of the estimator $\hat{\cE}_{h^*}$ under the optimal smoothing bandwidth is 
\begin{equation*}
\begin{aligned}
{\sf Var} \left(\hat{\cE}_{h^*}\right) 
 = \frac{1}{n}V_{\min} + \frac{1}{n^2}\left(m\cdot V_{q_0} + (n-m)\cdot O\left(\left(\frac{\log m}{m}\right)^{\frac{2}{d+4}}\right)\right).
\end{aligned}
\end{equation*}
\label{thm::NP_var}
\end{theorem}
Similar to Theorem~\ref{thm::P_var}, the variance in Theorem~\ref{thm::NP_var}
has three components:
the oracle variance $V_{\min}$, the excess variance due to the initial sampling density 
$\frac{1}{n^2}\cdot m\cdot V_{q_0} $,
and the excess variance from the estimator $\hat{r}_{h^*}(x)$.

To obtain the rate of the optimal allocation,
we equate the two excess variances:
\begin{align*}
&m\cdot V_{q_0} = (n-m)\cdot O\left(\left(\frac{\log m}{m}\right)^{\frac{2}{d+4}}\right)\\
\Longrightarrow \quad&m\cdot \left(\frac{m}{\log m}\right)^{\frac{2}{d+4}} \asymp n\\
\Longrightarrow \quad&m\asymp\left(\frac{n}{\log n}\right)^{\frac{d+4}{d+6}} \quad \mbox{(ignoring the $\log\log n$ and multi-logarithm terms)}.
\end{align*}
This choice of $m$ yields the following variance reduction rate.
\begin{corollary}
Assume (N1--2), (K1--2). 
When $m \asymp\left(\frac{n}{\log n}\right)^{\frac{d+4}{d+6}}$, and $h^*  \asymp\left(\frac{\log m}{m}\right)^{\frac{1}{d+4}}$, 
the variance of the estimator $\hat{\cE}_{h^*}$ is
\begin{align*}
{\sf Var} \left(\hat{\cE}_{h^*}\right) &= \frac{1}{n}V_{\min} + O\left(\frac{1}{n^2}\cdot n\cdot \log^{\frac{2}{d+4}} n\cdot\left(\frac{n}{\log n}\right)^{\frac{d+4}{d+6}\times \frac{-2}{d+4}}\right)\\
&= \frac{1}{n}V_{\min} \left(1 + O\left(\left(\frac{\log^{(4d+20)/(d+4)} n}{n}\right)^{\frac{2}{d+6}}\right)\right)\\
&= \frac{1}{n}V_{\min} \left(1 + O\left(\left(\frac{\log^5n}{n}\right)^{\frac{2}{d+6}}\right)\right)\\
\end{align*} 
\label{cor::mn}
\end{corollary}
Note that in the last equality in Corollary \ref{cor::mn},
we use the fact that $a_n = O(\log^{(4d+20)/(d+4)} n)$ implies $a_n = O(\log^5 n)$ to simplify the expression.
Corollary \ref{cor::mn} shows that under the optimal allocation, the excess variance
in the nonparametric importance sampler shrinks at rate
$O\left(\left(\frac{\log^5 n}{n}\right)^{\frac{2}{d+6}}\right)$.
When the dimension is small, say $d=1$ or $d=2$, the nonparametric method has 
an excess variance at rate $O\left(\left(\frac{\log^{5} n}{n}\right)^{\frac{2}{7}}\right)$
and $O\left(\left(\frac{\log^{5} n}{n}\right)^{\frac{1}{4}}\right)$,
which are just slightly slower than the rate of the parametric importance sampler under correct model
(the rate is $O\left(n^{-\frac{1}{3}}\right)$ by Corollary \ref{cor::Pmn}). 

Although the parametric method enjoys a fast convergence rate, it depends on a very 
restrictive assumption: the model has to be correctly specified.
This assumption is generally not true in most applications.
Thus, even if the nonparametric importance sampler has a slower variance reduction rate,
the nonparametric approach still has its own merit in applicability. 

{\color{black}There is a limitation in the nonparametric approach--the curse of dimensionality. When $d$ is not small, the convergence rate is very slow. So the nonparametric approach may not be applied to scenarios with more than $4$ variables. In this situation, even if the parametric model does not converge to the optimal sampler, it may still be a preferred approach. The parametric model will converge to the best possible result under the mis-specified model as long as the the conditions of Theorem~\ref{thm::Prate} hold.}

\begin{remark}
Note that the nonparametric rate can be improved if the regression function
is very smooth and we use a higher order kernel \citep{wasserman2006all}. 
When the regression function is in $\beta$-H\"older class with $\beta>2$,
we can boost convergence rate in Theorem~\ref{thm::NP_var} to $O\left(\left(\frac{\log m}{m}\right)^{\frac{\beta}{d+2\beta}}\right)$ 
and under the optimal allocation, the variance of nonparametric importance sampler will be
$$
\frac{1}{n}V_{\min} \left(1 + O\left(\left(\frac{\log^{d+2.5\beta} n}{n}\right)^{\frac{\beta}{d+3\beta}}\right)\right).
$$
When $\beta\rightarrow \infty$, the variance of nonparametric importance sampler becomes 
$\frac{1}{n}V_{\min} \left(1 + O\left(n^{-1/3}\right)\right)$ (ignoring the $\log n$ term), which 
recovers the parametric rate in Corollary~\ref{cor::Pmn}.
\end{remark}


\subsection{Oracle Properties}	\label{sec::thm::nonparametric}

In Lemma~\ref{lem::var}, we see that the oracle sampler $\hat{\cE}_{q^*}$ has the minimal variance.
Let 
$$
\mathcal{G} = \left\{f: \int f(x)dx = 1, f(x)=0\Rightarrow g(V(x))p(x)=0 \,\,\forall {x}\right\}
$$
be the collection of density functions that leads to an unbiased importance sampler
and let
$$
\Xi = \left\{\hat{\cE}_{q}: q\in\mathcal{G}\right\}
$$
be the collection of all possible unbiased importance samplers. 
Because all importance samplers from $\mathcal{G}$ is unbiased, any estimator $\hat{\cE}\in\Xi$ satisfies 
\begin{equation}
\E\left\|\hat{\cE}-\cE\right\|^2 = {\sf Var} \left(\hat{\cE}\right)
\label{eq::oracle0}
\end{equation}
so the oracle sampler $\hat{\cE}_{q^*}$ satisfies 
\begin{equation}
\E\left\|\hat{\cE}_{q^*}-\cE\right\|^2={\sf Var}\left(\hat{\cE}_{q^*}\right) = \inf_{\hat{\cE} \in \Xi} {\sf Var}\left(\hat{\cE}\right)
= \inf_{\hat{\cE} \in \Xi}\E\left\|\hat{\cE}-\cE\right\|^2.
\label{eq::oracle1}
\end{equation}
Because of equation \eqref{eq::oracle1}, $\hat{\cE}_{q^*}$ is called
the \emph{oracle for $\Xi$ with respect to the mean square error} in nonparametric theory;
see, e.g., page 60--61 in \cite{tsybakov2009introduction}.

We say $\hat{\cE}^\dagger\in\Xi$ \emph{satisfies the oracle inequalities}
if 
\begin{equation}
\frac{\E\left\|\hat{\cE}^\dagger-\cE\right\|^2}{\inf_{\hat{\cE} \in \Xi} \E\left\|\hat{\cE}-\cE\right\|^2} 
= \frac{\E\left\|\hat{\cE}^\dagger-\cE\right\|^2}{\E\left\|\hat{\cE}_{q^*}-\cE\right\|^2}
= 1 + o(1).
\end{equation}
Note that the estimators in the above expressions are all based on a size $n$ sample
and we do not include the subscript $n$ to abbreviation.
In nonparametric theory, 
an estimator with the oracle inequalities 
implies that the estimator is asymptotically as good as the optimal estimator.

The crude Monte Carlo sampler $\hat{\cE}_{p}$ (which samples only from the natural configuration density $p$ \citep{kroese:2011handbook}) obviously
does not satisfy the oracle inequalities because
$$
\frac{\E\left\|\hat{\cE}_p-\cE\right\|^2}{\inf_{\hat{\cE} \in \Xi}\E\left\|\hat{\cE}-\cE\right\|^2} 
= \frac{{\sf Var}\left(\hat{\cE}_{p}\right)}{{\sf Var}\left(\hat{\cE}_{q^*}\right)} = \frac{V_{\min} + V_{p}}{V_{\min}} = 1 +\frac{V_p}{V_{\min}}>1,
$$
where $V_p = \E(r(X^*))-\E^2\left(\sqrt{r(X^*)}\right)>0$.

The parametric importance sampler $\hat{\cE}_{\hat{\theta}_m}$ satisfies the oracle inequalities 
when the model is correctly specified 
(i.e. $r(x)=r_{\theta}(x)$ for some $\theta\in\Theta$).
To see this, 
recall Corollary \ref{cor::Pmn} and equation \eqref{eq::oracle0}:
$$
\frac{\E\left\|\hat{\cE}_{\hat{\theta}_m}-\cE\right\|^2}{\inf_{\hat{\cE} \in \Xi}\E\left\|\hat{\cE}-\cE\right\|^2}=\frac{{\sf Var} \left(\hat{\cE}_{\hat{\theta}_m}\right) }{{\sf Var}\left(\hat{\cE}_{q^*}\right)}= 1+ O\left( n^{-\frac{1}{3}}\right) = 1+o(1).
$$

However, when the model is incorrect, Theorem~\ref{thm::P_inconsistent} proves that
$\hat{\cE}_{\hat{\theta}_m}$ does not have the oracle inequalities:
$$
\frac{\E\left\|\hat{\cE}_{\hat{\theta}_m}-\cE\right\|^2}{\inf_{\hat{\cE} \in \Xi}\E\left\|\hat{\cE}-\cE\right\|^2}=\frac{{\sf Var} \left(\hat{\cE}_{\hat{\theta}_m}\right) }{{\sf Var}\left(\hat{\cE}_{q^*}\right)} = 1+ \frac{V_{\theta^*}}{V_{\min}} > 1.
$$

The nonparametric importance sampler has a good advantage that it satisfies the oracle inequalities in
most cases.
By Corollary \ref{cor::mn} and equation \eqref{eq::oracle0}, 
$$
\frac{\E\left\|\hat{\cE}_{h^*}-\cE\right\|^2}{\inf_{\hat{\cE} \in \Xi}\E\left\|\hat{\cE}-\cE\right\|^2}=\frac{{\sf Var} \left(\hat{\cE}_{h^*}\right) }{{\sf Var}\left(\hat{\cE}_{q^*}\right)}=
1 + O\left(\left(\frac{\log n}{n}\right)^{\frac{2}{d+6}}\right) =1+ o(1).
$$
Thus, without any further information about the structure of $r(x)$, we recommend to use the nonparametric importance sampler
since it behaves asymptotically as good as the oracle (optimal) importance sampler.

\begin{remark}
How we obtain the oracle property is very different from 
the classical approach.
Many estimators with oracle properties are constructed by minimizing
an estimated risk \citep{tsybakov2009introduction}.
That is, for a collection of estimators,  the risk of each of them is estimated and the one that minimizes the (estimated) risk is chosen. 
When the risk is consistently estimated uniformly for all estimators, 
this procedure leads to an estimator with the oracle property.
However, in our case, we do not consider any risk estimator nor do we choose an estimator from
many possible candidates, but we still obtain the oracle property.
\end{remark}


\section{Empirical Analysis}\label{sec::emp}

To evaluate the empirical performances of the importance samplers, this section presents an implementation guideline, a numerical study, and a case study. 

\subsection{Implementation Guideline}
To implement parametric or nonparametric importance sampling, we can follow the procedure in Figure~\ref{fig::parametric} or Figure~\ref{fig::nonparametric}, respectively. In practice, $n$ is typically determined based on the available computational budget. We can choose $m$ according to the optimal allocation rate, 
$m \asymp n^{\frac{2}{3}}$ 
in Corollary~\ref{cor::Pmn} for parametric importance sampler, or 
$m \asymp\left(\frac{n}{\log n}\right)^{\frac{d+4}{d+6}}$ 
in Corollary~\ref{cor::mn} for nonparametric importance sampler. In the range of experiments we present below, any choice of multiplicative constant between two and six results in similar empirical performances of the importance samplers. 


{\color{black} In the first stage, we can simply choose the natural configuration density $p$ as the initial sampling density $q_0$ to maintain the same error rate as the crude Monte Carlo sampler. In the second stage, once} we build a regression model $\hat{r}(x)$ for the unknown conditional expectation $r(x) = \E(g^2(V(X))|X=x)$, we can exactly sample from  $q(x) \propto \sqrt{\hat{r}(x)} \cdot p(x)$ using the acceptance-rejection method \citep[][p.59]{kroese:2011handbook} {\color{black} by using $p(x)$ as the proposal (or envelope) density and finding an upper bound on $\sqrt{\hat{r}(x)}$. The upper bound can be computed numerically or even known from physical/engineering knowledge of the system (e.g., the simulation output $V$ can take up to a certain value due to a physical/engineering limit, or $r(x)$ is a probability). As an alternative to the acceptance-rejection method, Markov chain Monte Carlo methods can be used as well. 
}  

Also, for an importance sampling estimator (e.g., in \eqref{eq:param_IS_estimator} or \eqref{eq:nonparam_IS_estimator}), the normalization constant of $q(x)$ can be calculated to a desired level of accuracy{\color{black}, independent of $n$,} by using a numerical integration such as quadrature for low-dimensional $x$ and Monte Carlo integration for high-dimensional $x$. {\color{black}Thus, we can ensure that estimation of the normalization constant will not affect the asymptotic optimality and empirical
performance of the importance sampling estimator. If desired, one can avoid calculating the normalization constant by using a self-normalized importance sampling estimator at the expense of bias. More in-depth discussion of this trade-off is made by \citet{owen2018mcbook}.}

Note that in practice, the computational costs of the acceptance-rejection method and the numerical integration are negligible (a matter of hours, if not minutes) compared to running high-fidelity simulation models such as those discussed in Section~\ref{sec::intro} and the simulation model in our case study. For example, a simulation experiment can take days (if not weeks) even with a supercomputer \citep{Manuel2013,graf2016high}.

\subsection{Numerical Study} \label{sec::emp::num_study}
Our numerical study considers two examples, one with normal distributions for $X$ and $V|X$ and the other with exponential distributions for $X$ and $V|X$. Motivated by our case study, we estimate the probability $\cE =  P(V >\xi ) = \mathbb{E}(g(V(X)))$ for $g(V) = 1(V> \xi)$ and a pre-specified $\xi > 0$. {\color{black}As a baseline, w}e set $\xi$ such that $P(V >\xi )$ is equal to $0.5$ {\color{black}(unless specified otherwise)}  regardless of the input configuration dimension $d$ because it is known that the performance of importance sampler often depends on the probability being estimated  \citep{heidelberger1995fast,kroese:2011handbook}.

We vary the total sample size $n = 1000, 2000, 4000, 8000$ and the input configuration dimension $d = 1,2, 4$ to see their impacts on the mean squared error (MSE) 
\begin{equation*}
\textrm{MSE} = \frac{1}{n_{MC}} \sum_{i=1}^{n_{MC}} \left\|\hat{\cE}_i-\cE\right\|^2 ,
\end{equation*}
where $\hat{\cE}_i$ is an estimate of the $i$th replication and the total number of Monte Carlo replications, $n_{MC}$, is set as 10,000 to obtain reliable results. We use high-performance computing (Lenovo NextScale E5-2680 v4, total 112 cores with 1TB RAM) for our simulation experiments, and they take several weeks in total. The R scripts for the experiments are available as a supplementary material. 

We consider two parametric importance samplers, one with a correct model of 
\begin{align*}
r(x) &= \E(g^2(V(X))|X=x) = \E(1(V(X) > \xi)|X=x) = P(V >\xi \mid X =x ) 
\end{align*}
 and the other with an incorrect model, and a nonparametric importance sampler. To build the parametric models of $r(x)$, we use the sample of size $m = \left\lceil 2n^{\frac{2}{3}} \right\rceil$. 
 For the nonparametric model, we use $m = \left\lceil 6\left(\frac{n}{\log n}\right)^{\frac{d+4}{d+6}} \right\rceil$. 

To sample from the importance sampling density $q(x) \propto \sqrt{\hat{r}(x)} \cdot p(x)$ using the acceptance-rejection method, we use $p(x)$ as the envelope density because $p(x) \ge \sqrt{\hat{r}(x)}p(x)$  
in the examples: We sample $x$ from $p(x)$ and accept $x$ with the probability $\sqrt{\hat{r}(x)}$.   To compute the normalizing constant of $q(x)$, we use Monte Carlo integration. Since we know the true $r(x)$, which is unknown in practice, in the examples, we calculate the true $\cE =  P(V >\xi )$ and $V_{\min}$ using Monte Carlo integration and use them to calculate MSEs and demonstrate how empirical results conform to the theoretical predictions made in Section~\ref{sec::thm}. 


\subsubsection{Example 1: normal-normal data generating model}	\label{sec::ex1}
As a modification of an example in \citet{ackley1987}, we use the data generating model where the $d$-dimensional input vector $X = (X^{(1)}, \ldots, X^{(d)})'$ follows a multivariate normal distribution with zero mean and identity covariance matrix, and the output $V$ at $X$ follows $N(\mu(X), 1)$ with 
\begin{align*}
\mu(X) &=20 \left( 1-\exp \left(-0.2 \sqrt{\frac{1}{d} \|{X}\|^2 } \right)  \right) + \left(\exp\left(1\right) - \exp \left(\frac{1}{d}\sum_{i=1}^{d} \cos(2\pi  X^{(i)})\right)\right) .
\end{align*}
Thus, we have
\begin{align*}
r(x) 
= P(V >\xi \mid X =x ) 
= 1-\Phi(\xi- \mu(x)) ,
\end{align*}
where $\Phi(\cdot)$ is the CDF of a standard normal distribution. As parametric models of $r(x)$, we consider two models:
\begin{itemize}
	\item[(i)] Correct model: We use $r_\theta(x) = 1-\Phi(\xi- \hat{\mu}(x))$, where 
	\begin{align*}
	\hat{\mu}(x) &=20 \left( \theta_0 -\exp \left(-0.2 \sqrt{\frac{1}{d} \sum_{i=1}^{d} \theta_i^2 (x^{(i)})^2 } \right)  \right) + \left(\theta_0\exp\left(1\right) - \exp \left(\frac{1}{d}\sum_{i=1}^{d} \theta_i\cos(2\pi  x^{(i)})\right)\right) ,
	\end{align*}
	such that $r_\theta(x) = r(x)$ for some $\theta = (\theta_0, \ldots, \theta_d)' \in\Theta$. For fitting with a least square method, the initial parameters are set at the correct values, i.e., $\hat{\theta}_0=\ldots=\hat{\theta}_d=1$, in the implementation.
	\item[(ii)] Incorrect model: We use the logistic regression model $r_\theta(x) = \left( 1 + e^{\theta_0 + \theta_1 x^{(1)} + \ldots + \theta_d x^{(d)}} \right)^{-1}$ such that $r_\theta(x) \neq r(x)$ for all $\theta\in\Theta$.  
	For least square fitting, the initial parameters are set at $\hat{\theta}_0=\ldots=\hat{\theta}_d=0$.
\end{itemize}
As a nonparametric model of $r(x)$, we use the kernel regression model $\hat{r}_h(x)$ with the Gaussian kernel and the smoothing bandwidth $h$ chosen by cross-validation.



\begin{figure}[!htb] 
	\centering
	{\centering 
		\subfigure[Varying $n$ with fixed $d=1$.]{ \includegraphics[width=0.46\linewidth]{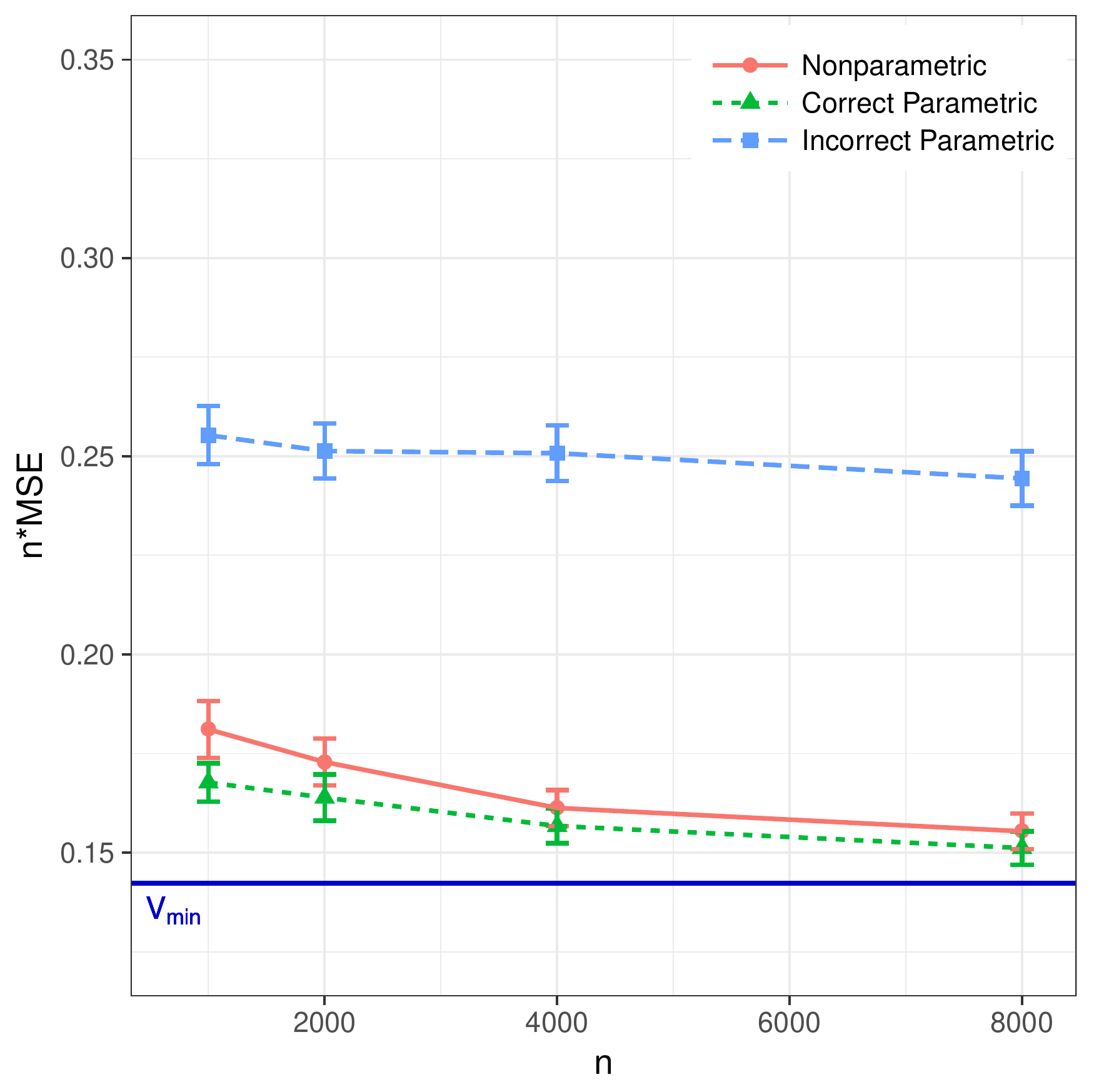}}
		\subfigure[Varying $n$ with fixed $d=2$.]{ \includegraphics[width=0.46\linewidth]{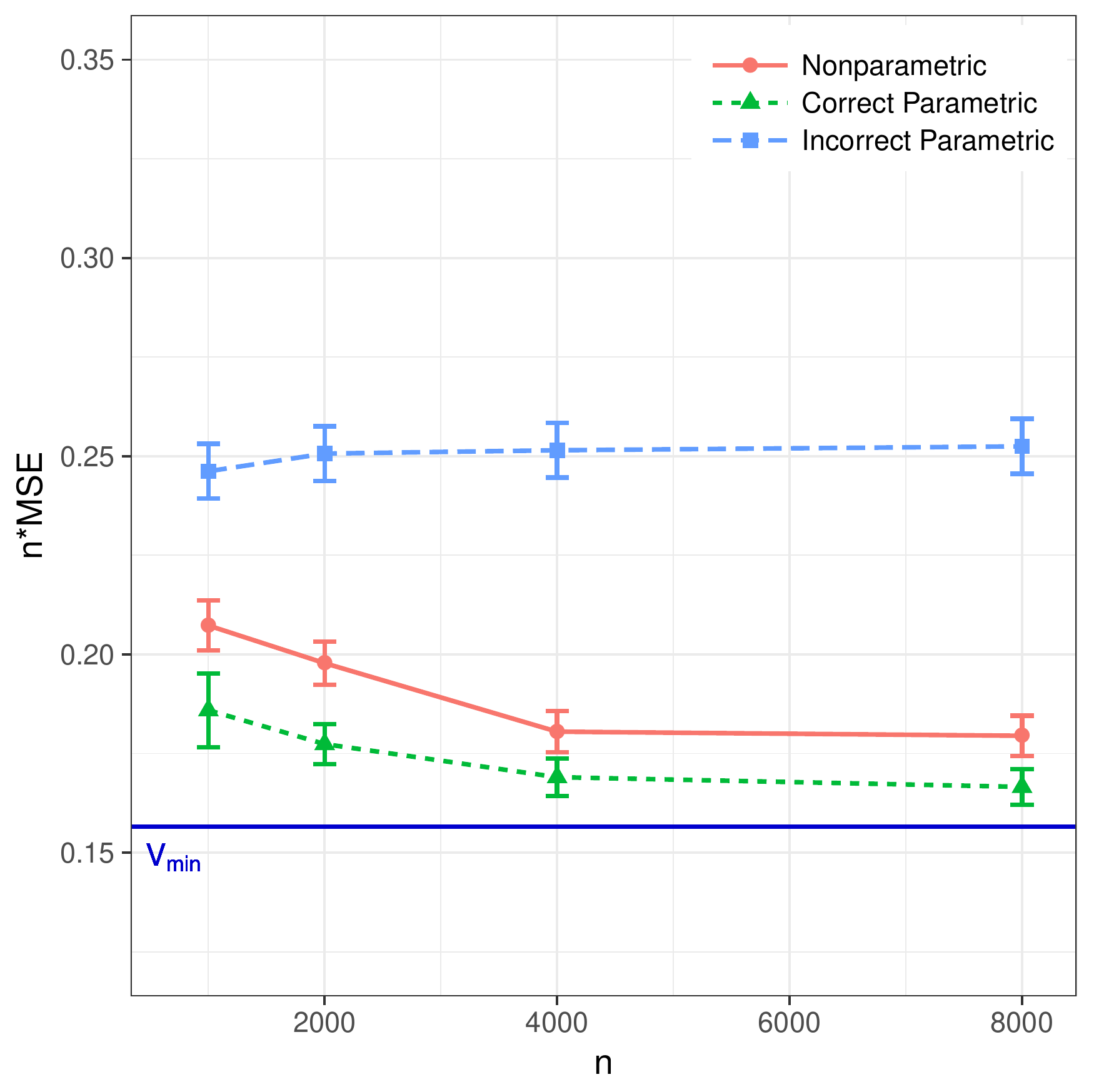}}
		\subfigure[Varying $n$ with fixed $d=4$.]{ \includegraphics[width=0.46\linewidth]{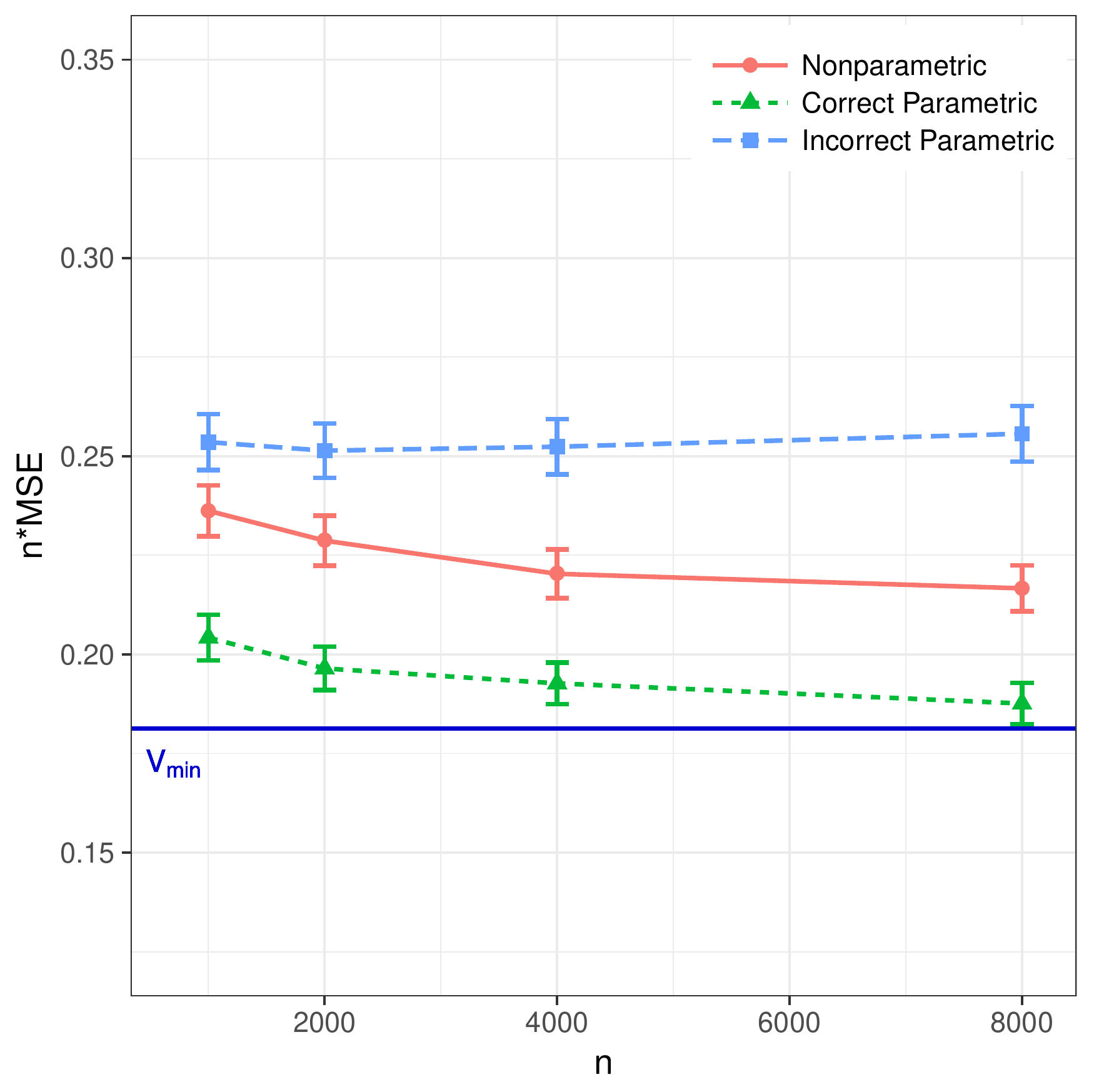}}
		\subfigure[Varying $d$ with fixed $n=8000$.]{ \includegraphics[width=0.46\linewidth]{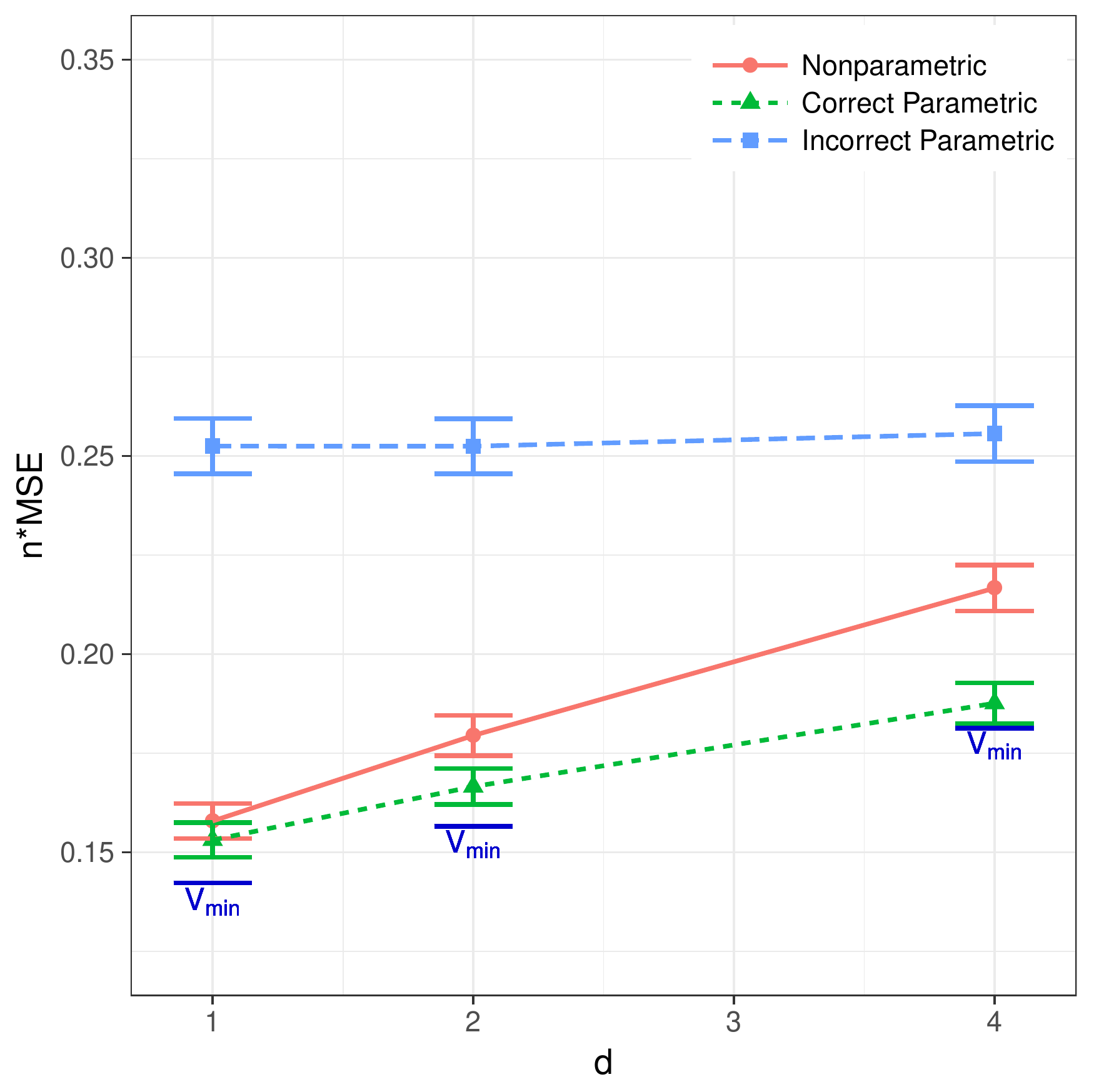}}
	} 
	\caption{For the normal-normal data generating model, we compare the three importance samplers in terms of their scaled estimation error, $n\textrm{MSE}$, against the total sample size $n$ for the input dimension $d = 1$ in (a), $d = 2$ in (b), and $d = 4$ in (c). While fixing $n = 8000$, $d$ is varied in (d). An error bar represents the 95\% confidence interval based on the Monte Carlo error with 10,000 replications.} \label{fig:norm-norm-results} 
\end{figure}

Figures~\ref{fig:norm-norm-results}(a)--\ref{fig:norm-norm-results}(c) show how $n \textrm{MSE}$ varies as $n$ increases for $d=1,2,4$.  For the correct parametric model, $n{\sf Var} \left(\hat{\cE}_{\hat{\theta}_m}\right)$ is predicted to be $V_{\min} \left(1+ O\left( n^{-\frac{1}{3}}\right)\right)$ by Corollary~\ref{cor::Pmn}. Recalling that MSE is equal to the variance of an importance sampler, because of its unbiasedness, we see that $n\textrm{MSE}$ approaches $V_{\min}$ as $n$ increases, with roughly the same rate regardless of $d$. 
For the incorrect parametric model, as foreseen by Theorem~\ref{thm::P_inconsistent}, $n\textrm{MSE}$ fails to approach $V_{\min}$ as $n$ increases, because $n {\sf Var} \left(\hat{\cE}_{\hat{\theta}_m}\right) 
\geq  V_{\min} (1+ O(1))$. As anticipated by Corollary~\ref{cor::mn}, $n\textrm{MSE}$ for the nonparametric model approaches $V_{\min}$ as $n$ increases, with an apparently slower rate for larger $d$, because $n{\sf Var} \left(\hat{\cE}_{h^*}\right) = V_{\min} \left(1 + O\left(\left(\frac{\log n}{n}\right)^{\frac{2}{d+6}}\right)\right)$.

Figure~\ref{fig:norm-norm-results}(d) shows $n\textrm{MSE}$ against $d$ for fixed $n = 8000$.  Regardless of $d$, $n\textrm{MSE}$ of the correct parametric importance sampler stays close to $V_{\min}$. In contrast, $n\textrm{MSE}$  for the incorrect parametric importance sampler essentially remains the same as $d$ varies in this example, although this observation cannot be taken as a general pattern because the input configuration dimension $d$ impacts how \emph{incorrect} the model is. While the nonparametric importance sampler performs almost as well as the correct parametric importance sampler when $d=1$, the performance gap widens as $d$ increases since $n$ is fixed.

If $X$ was sampled only from the natural configuration density $p$ instead of an importance sampling density $q$ in the estimator in \eqref{eq::est1}, then this simple baseline approach, commonly called crude Monte Carlo (CMC) \citep{kroese:2011handbook}, results in the estimator having the theoretical $n\textrm{MSE}$ of 0.25 $(=P(V >\xi )(1-P(V >\xi ))$. In this example, the incorrect parametric importance sampler essentially does not improve over the baseline.

As $d$ increases, $V_{\min}$ approaches the baseline $n\textrm{MSE}$ of 0.25 in Figure~\ref{fig:norm-norm-results}(d). 
This increasing inefficiency of optimal importance sampling with respect to $d$ is regarded as peculiar to this example, because $V_{\min}$ in \eqref{eq:V_min} depends on $d$ only through $g(V(X))$. Thus, this observation should not be interpreted as a manifestation of curse of dimensionality known in the importance sampling literature \citep[e.g.,][]{au2003}, which may occur when the \emph{approximation} of optimal density $q^*$ becomes harder as $d$ increases. In contrast, it is known that the optimal importance sampler theoretically attains $V_{\min}$ of zero regardless of $d$ for deterministic simulation models with any nonnegative function $g(v)$ \citep{kahn1953}. 

\begin{figure}[!htb] 
	\centering{\color{black}
	{\centering 
		\subfigure[$\textrm{RMSE}/\cE$ vs. $\cE =  P(V >\xi )$ in log scale.]{ \includegraphics[width=0.46\linewidth]{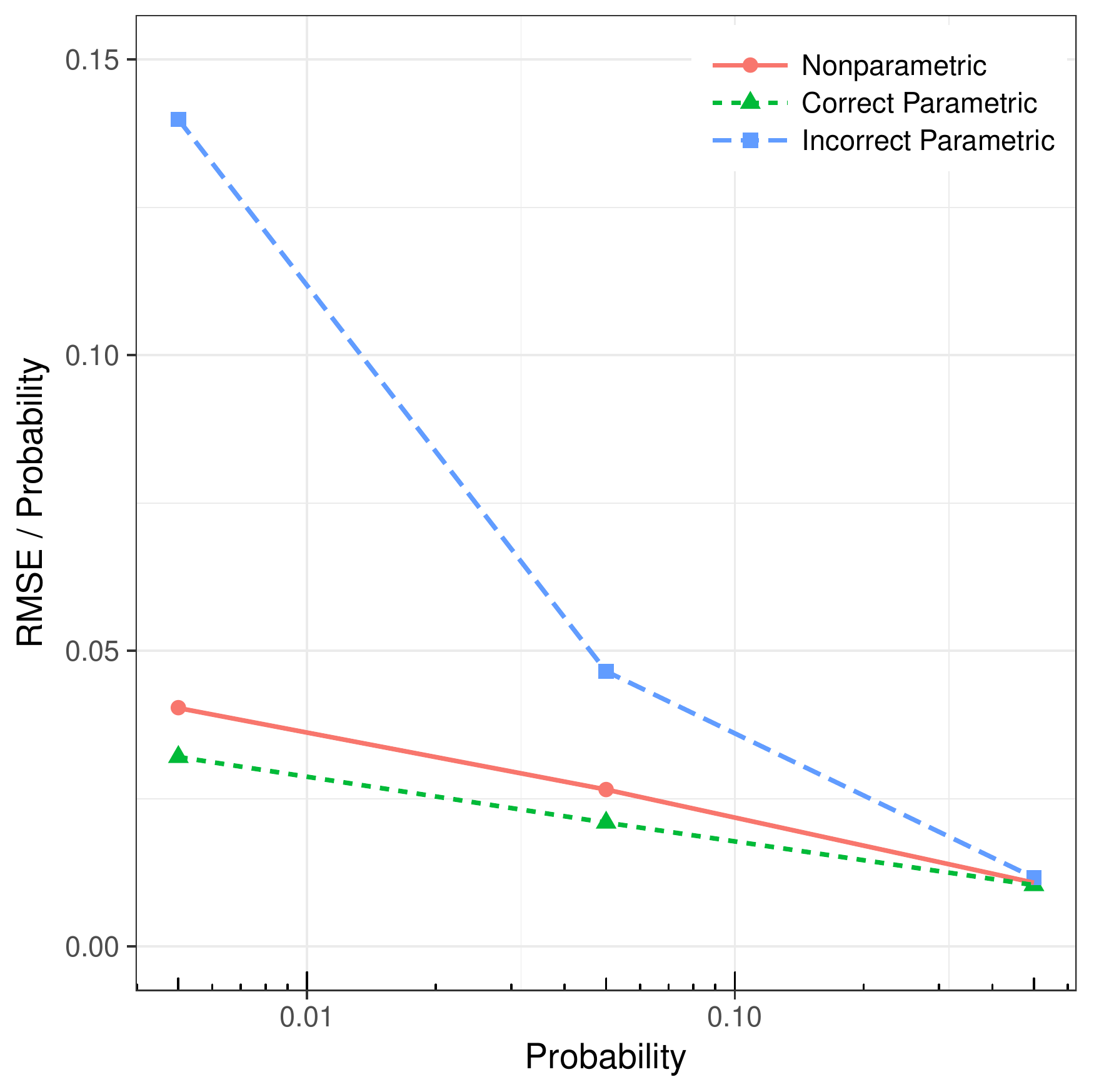}}
		\subfigure[Computational saving vs. $\cE =  P(V >\xi )$ in log scale.]{ \includegraphics[width=0.46\linewidth]{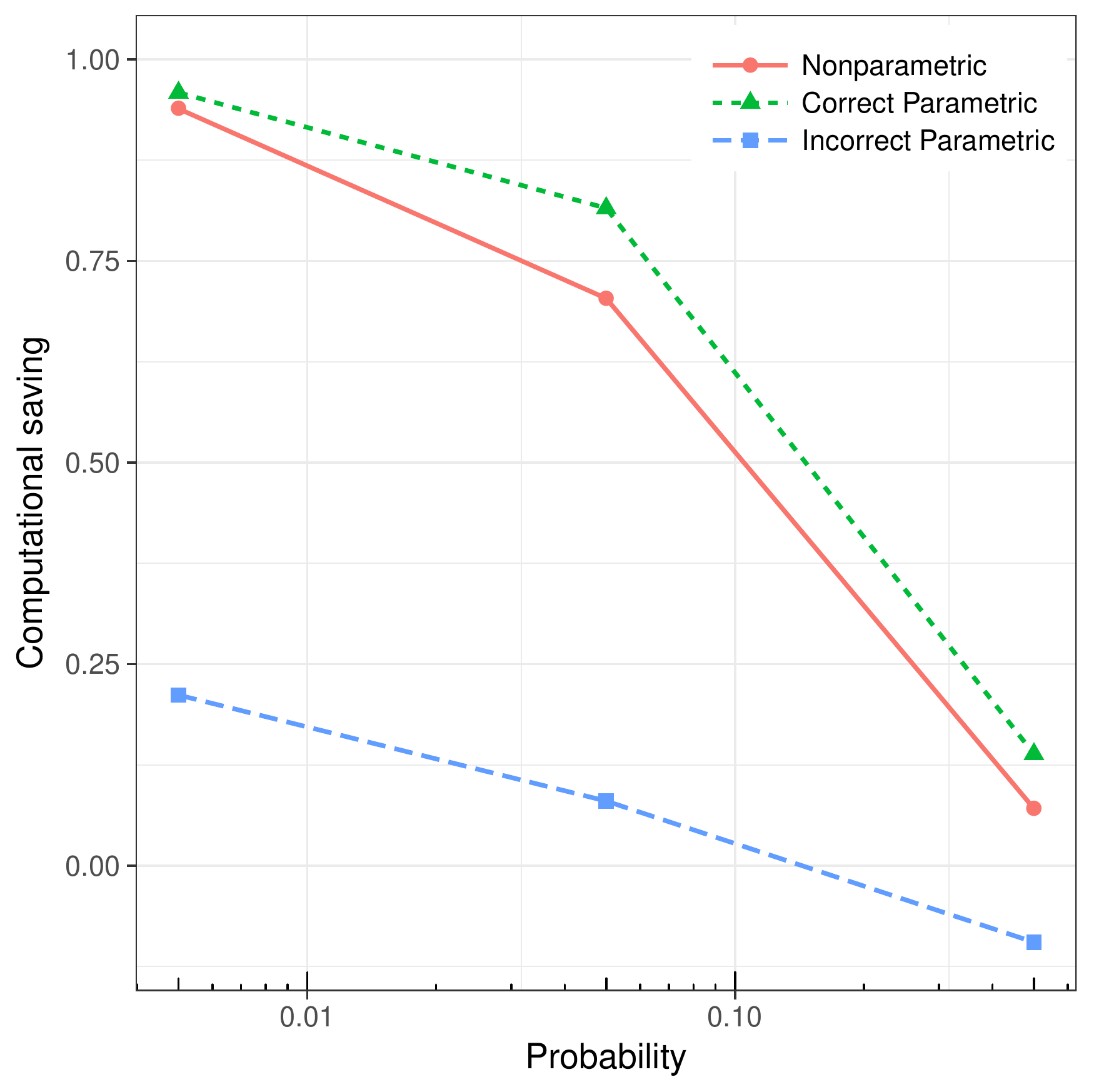}}
	}
	\caption{For the normal-normal data generating model, we compare the three importance samplers by varying the estimand (or the probability being estimated) $\cE =  P(V >\xi ) = 0.005, 0.05, 0.5$ while fixing the total sample size $n = 8000$ and the input dimension $d = 1$. We examine the normalized root mean squared error (i.e., $\textrm{RMSE}/\cE$) in (a) and computational saving against CMC in (b). The computational saving is defined as $(n_{(CMC)} - n ) / n_{(CMC)}$, where $n_{(CMC)}= \cE(1-\cE)/(\textrm{Standard error})^2$ is the theoretical sample size required for CMC to attain the same standard error as the importance sampler. The experiment results are averaged across 10,000 Monte Carlo replications. For the nonparametric importance sampler when $\cE = 0.05$, one estimate (out of 10,000) of $\cE$ is observed to be greater than one and thus excluded in the analysis.}\label{fig:norm-norm-varying-p}} 
\end{figure}

{\color{black}We also vary the estimand $\cE =  P(V >\xi ) = 0.005, 0.05, 0.5$ to examine its impact on the performance of the importance samplers, while fixing $n = 8000$ and $d = 1$. The initial sampling density $q_0$ is set as the uniform distribution on $(-5,5)$ to broadly cover the support where rare events of interest can happen. Figure~\ref{fig:norm-norm-varying-p}(a) suggests that as the estimand decreases by a factor of 10 (i.e., 0.5, 0.05, 0.005), the normalized root mean squared error (i.e. $\sqrt{\textrm{MSE}}/\cE$) increases more slowly for the correct parametric and nonparametric importance samplers than the incorrect parametric importance sampler.  Figure~\ref{fig:norm-norm-varying-p}(b) shows the well-known phenomenon in the literature that importance samplers save more against CMC as the estimand is the probability of a rarer event. The savings of the correct parametric and nonparametric importance samplers quickly go over 90\% as $\cE$ decreases to 0.005.}



\subsubsection{Example 2: exponential--exponential data generating model}	\label{sec::ex2}
Here, we consider a data generating model where both $X$ and $V|X$ follow exponential distributions that have heavier tails than normal distributions and allow analytical calculations of key objects of interest such as the estimand $\cE$, the conditional expectation $r(x) = \E(g^2(V(X))|X=x)$,  the optimal sampling density $q^*(x)$, and the oracle variance $V_{\min}$. 

Let $X = (X^{(1)}, \ldots, X^{(d)})'$ be a vector of $d$ independent exponential random variables with the identical mean $1/\lambda > 0$ so that the natural configuration density 
$$
p(x) = \lambda^d e^{-\lambda \left(x^{(1)}+ \ldots + x^{(d)}  \right)} .
$$
Given a configuration $X$, let $V$ follow an exponential distribution with a mean $1/\left(X^{(1)}+ \ldots + X^{(d)}\right)$. 
In our simulation experiment, we fix $\lambda = 1$.

With the given data generating model, we can analytically calculate 
$$\cE = P(V >\xi ) = \left(\frac{\lambda}{\xi + \lambda}\right)^d,$$ 
$$r(x) = \E(g^2(V(X))|X=x) = P(V >\xi \mid X =x ) = e^{-\xi\left(x^{(1)}+ \ldots + x^{(d)}\right)},$$ and
$$q^*(x) \propto \sqrt{r(x)} \cdot p(x) \propto e^{-\left(\frac{\xi}{2} + \lambda \right) \left(x^{(1)}+ \ldots + x^{(d)}  \right)},$$ 
which implies that $q^*$ is the joint density of $d$ independent exponential random variables with the identical mean $1/\left(\xi/2 + \lambda \right)$. We determine
\begin{align*}
\xi &= \frac{\lambda}{\left[P\!\left(V >\xi \right)\right]^{(1/d)}} - \lambda
\end{align*}
by plugging $P\!\left(V >\xi \right) = 0.5$.  We also know  
$$r^\dagger(x) = \E(g(V(X))|X=x) = P(V >\xi \mid X =x )$$
and calculate 
\begin{align*}
V_{\min} &= \mathbb{E}^2\left(\sqrt{r(X)}\right)-\mathbb{E}^2(r^\dagger(X))
\\&=  \left(\frac{\lambda}{\xi/2 + \lambda}\right)^{2d} - \left(\frac{\lambda}{\xi + \lambda}\right)^{2d} .
\end{align*}

Similar to the normal-normal example, we consider two parametric models of $r(x)$:
\begin{itemize}
	\item[(i)] Correct model: We use $r_\theta(x) = e^{\theta_0 + \theta_1 x^{(1)} + \ldots + \theta_d x^{(d)}}$.  For least square fitting, the initial parameters are set at $\hat{\theta}_0=\ldots=\hat{\theta}_d=0$. We set $r_{\hat{\theta}}(x)  = 1$ if $r_{\hat{\theta}}(x) > 1$. 
	\item[(ii)] Incorrect model: We use the logistic regression model, $r_\theta(x) = \left( 1 + e^{\theta_0 + \theta_1 x^{(1)} + \ldots + \theta_d x^{(d)}} \right)^{-1}$, with the initial parameters $\hat{\theta}_0=\ldots=\hat{\theta}_d=0$ for least square fitting.  
\end{itemize}
As a nonparametric model, we use the kernel regression model $\hat{r}_h(x)$ as in the normal-normal example. 

\begin{figure}[!htb] 
	\centering
	{\centering 
		\subfigure[$n\textrm{MSE}$ (calculated after removing nonparametric estimates greater than one) vs. $n$ for $d=1$.]{ \includegraphics[width=0.46\linewidth]{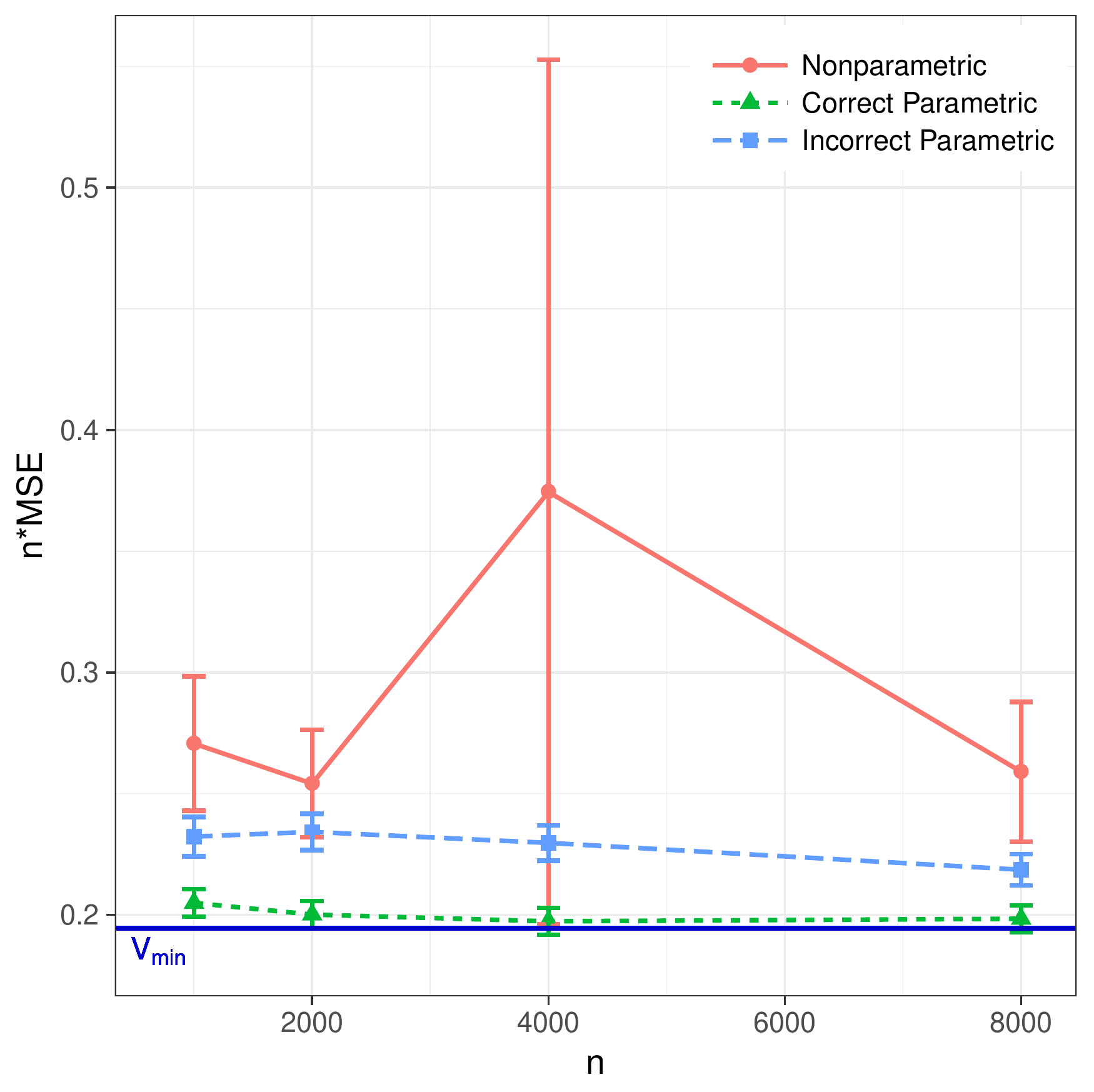}}
		\subfigure[Nonparametric estimate $\hat{\cE}_h$  (outliers greater than one are presented as numbers at the top) vs. $n$ for $d=1$.]{ \includegraphics[width=0.51\linewidth, trim={0 0.2in 0.3in 0},clip]{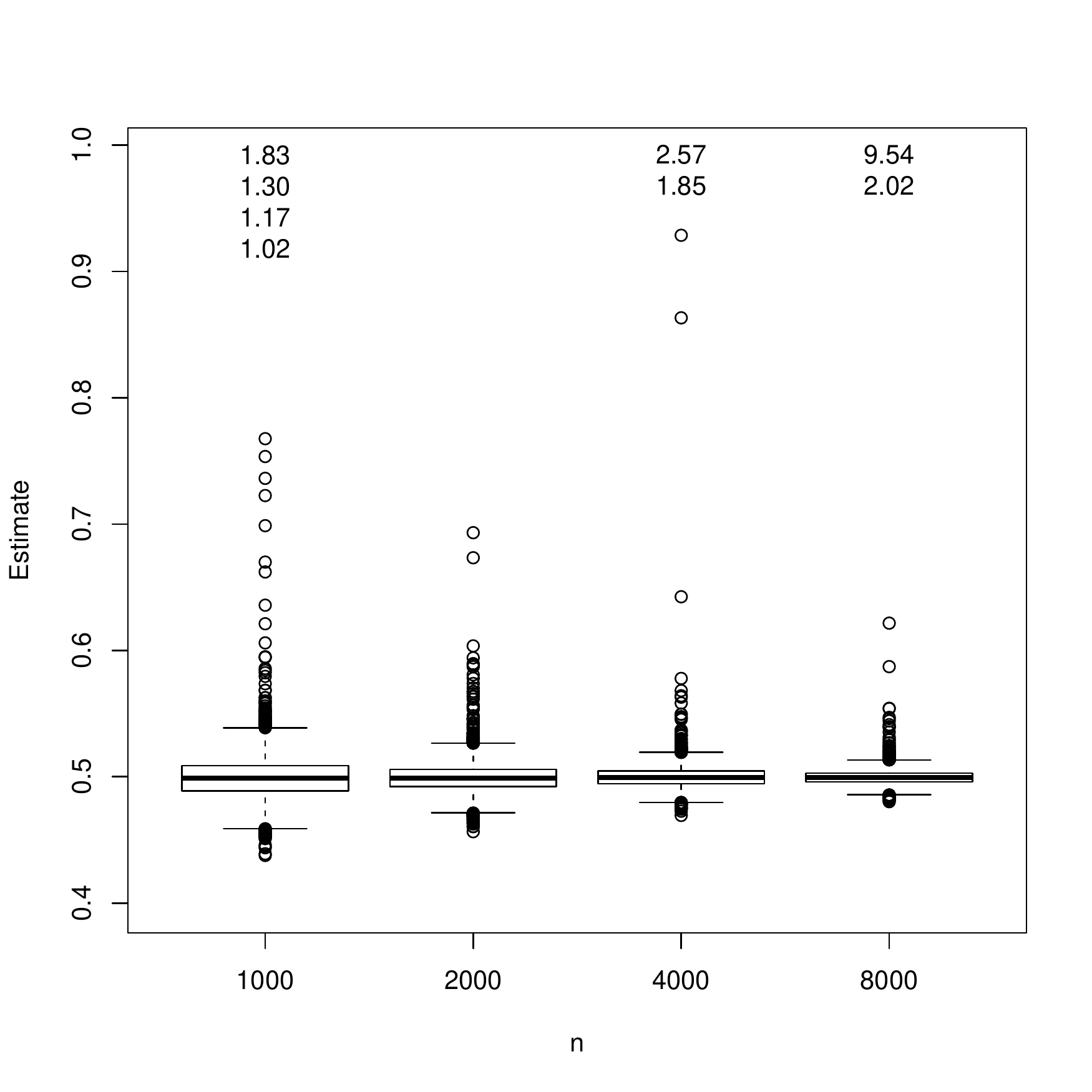}}
	}
	\caption{For the exponential-exponential data generating model,  we compare the three importance samplers in terms of $n\textrm{MSE}$ against $n$ for $d = 1$ in (a), where an error bar represents the 95\% confidence interval based on the Monte Carlo error with 10,000 replications. In (b), the Tukey box plots are drawn based on the 10,000 nonparametric estimates for each $n$. The ends of the whiskers represent the most extreme data points which are not exceeding 1.5 times the interquartile range from the box.} \label{fig:exp-exp-results} 
\end{figure}

Figure~\ref{fig:exp-exp-results}(a) plots $n \textrm{MSE}$ versus $n$ for $d=1$ with respect to the three importance samplers. The behaviors of correct and incorrect parametric importance samplers echo what we see in the normal-normal example. In contrast, the nonparametric importance sampler behaves irregularly (note that $n\textrm{MSE}$ in Figure~\ref{fig:exp-exp-results}(a) is calculated after discarding the nonparametric estimates exceeding one). Figure~\ref{fig:exp-exp-results}(b) shows the Tukey box plots of the nonparametric estimates over different $n$. The interquartile range (i.e., box height) decreases as $n$ increases (note that the interquartile range of estimates may be comparable to the square root of MSE, but not directly to $n\textrm{MSE}$ in Figure~\ref{fig:exp-exp-results}(a)). The magnitudes of outliers (e.g., estimates greater than one, presented as numbers at the top of Figure~\ref{fig:exp-exp-results}(b) for each $n$) suggest that the sampling distribution of the nonparametric estimator might be heavy-tailed for this example.

\begin{figure}[!htb] 
	\centering{\color{black}
	{\centering 
		\subfigure[$\widehat{\textrm{ESS}}$.]{ \includegraphics[width=0.46\linewidth]{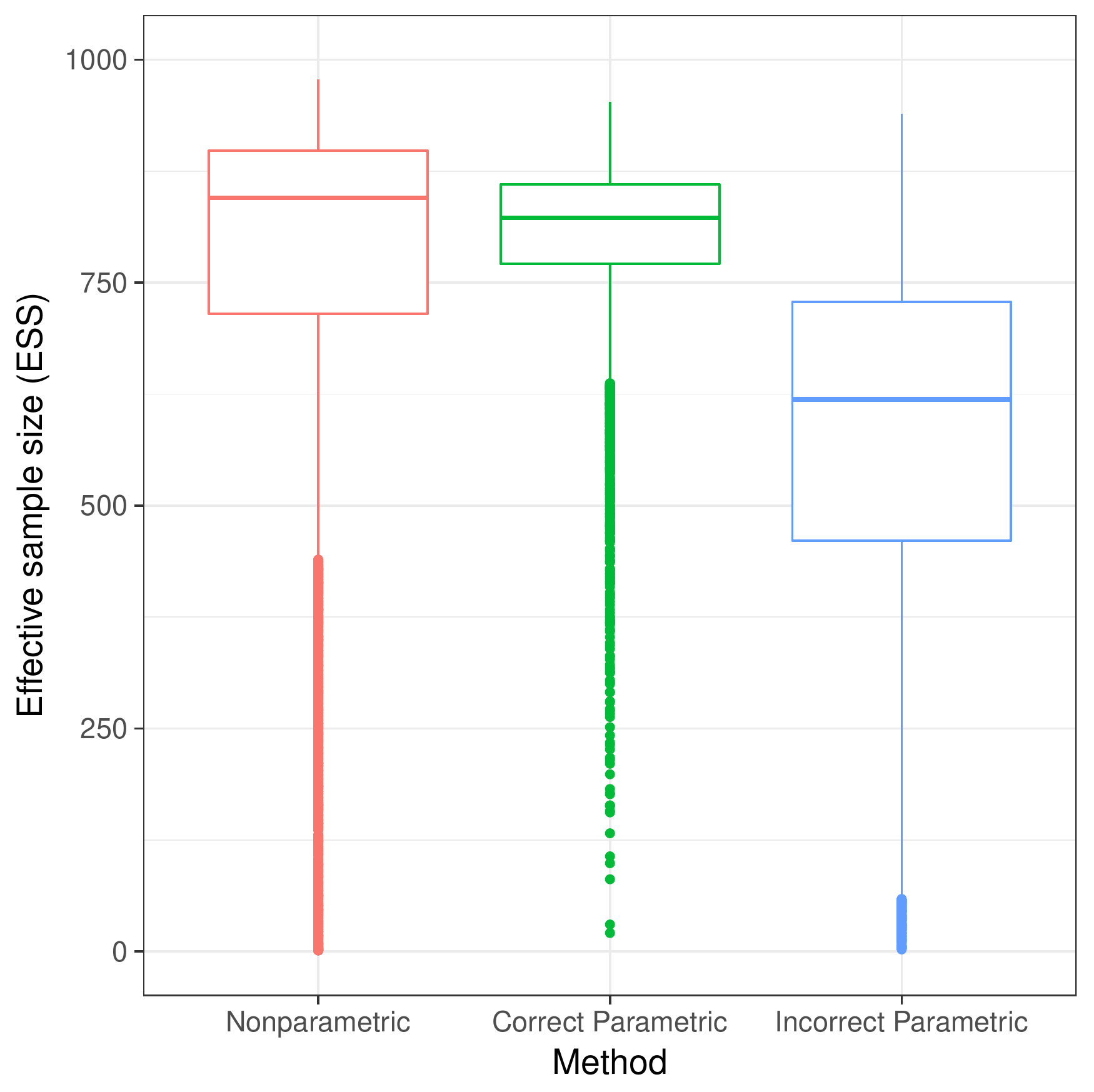}}
		\subfigure[$\widetilde{\textrm{ESS}}(g)$.]{ \includegraphics[width=0.46\linewidth]{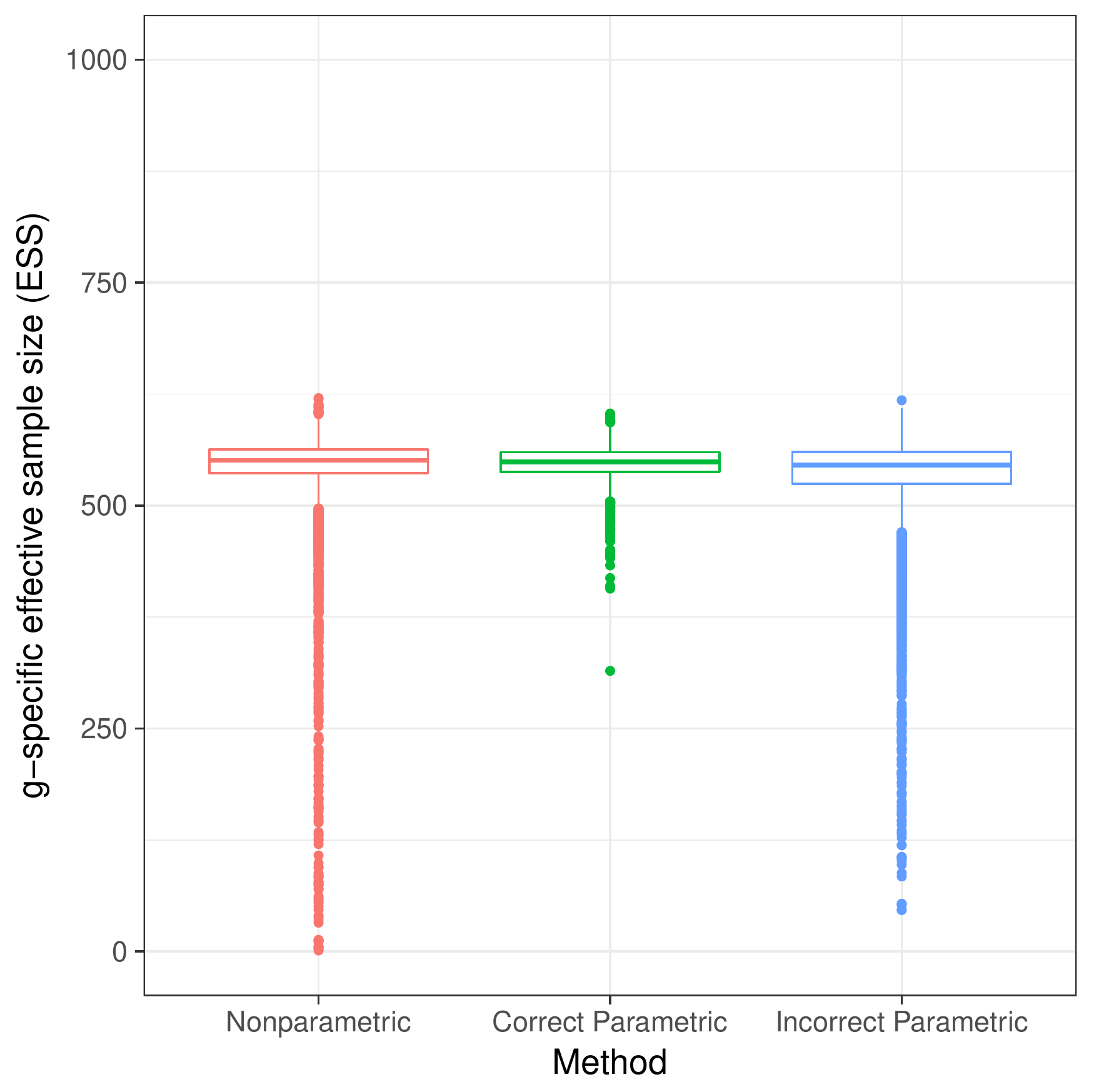}}
	}
	\caption{For the exponential-exponential data generating model,  we compare the three importance samplers in terms of their effective sample size (ESS), $\widehat{\textrm{ESS}}$, in \eqref{eq:ESS} and $g$-specific ESS, $\widetilde{\textrm{ESS}}(g)$, in \eqref{eq:ESS_g} when $d = 1$ and $n=1000$. The box plots are created based on 10,000 replications. The nonparametric importance sampler yields $\widetilde{\textrm{ESS}}(g)$ close to one, explaining the erratic $n\textrm{MSE}$ observed in Figure~\ref{fig:exp-exp-results}(a).}\label{fig:exp-exp-ESS}}  
\end{figure}

{\color{black}We calculate effective sample sizes (ESSs) as diagnostic tools of importance sampler performance, where a too small ESS compared to $n$ is interpreted as a sign that the importance sampling is unreliable. An ESS widely used in the importance sampling literature \citep{kong1992note,elvira2018rethinking} is
\begin{align}
\widehat{\textrm{ESS}} = \frac{\left(\sum_{i=1}^n w_i \right)^2}{\sum_{i=1}^n w_i^2} , \label{eq:ESS}
\end{align}
where $w_i$ is the likelihood ratio or importance weight for $X_i$. In the first stage ($i=1,\ldots,m$), both parametric and nonparametric importance samplers have $w_i = \frac{p(X_i)}{q_0(X_i)}$. In the second stage ($i = m+1, \ldots, n$), the parametric importance sampler has $w_i = \frac{p(X_i)}{q^*_{\hat{\theta}_m}(X_i)}$ and the nonparametric importance sampler has $w_i = \frac{p(X_i)}{\hat{q}^*_h(X_i)}$. While $\widehat{\textrm{ESS}}$ in \eqref{eq:ESS} can be conveniently used for any simulation model or $g(V)$, the following ESS specific to $g(V)$ can be a better diagnostic of importance samplers, as demonstrated for deterministic simulation models (where $V(X)$ is a deterministic function of $X$) \citep{evans1995methods,owen2018mcbook}:
\begin{align}
\widetilde{\textrm{ESS}}(g) = \frac{\left(\sum_{i=1}^n \tilde{w}_i(g) \right)^2}{\sum_{i=1}^n \tilde{w}_i(g)^2} , \label{eq:ESS_g}
\end{align}
where $\tilde{w}_i(g) = \left| g(V) \right| w_i$. For stochastic simulation models, as illustrated in Figure~\ref{fig:exp-exp-ESS}(a), a very small $\widehat{\textrm{ESS}}$ (close to one) does not necessarily indicate a poor estimate because correct parametric importance samplers always yield good estimates of $\cE$. On the other hand, $\widetilde{\textrm{ESS}}(g)$ is shown to be an effective diagnostic; in Figure~\ref{fig:exp-exp-ESS}(b), it is confirmed that when $\widetilde{\textrm{ESS}}(g)$ is close to one, the corresponding estimates of nonparametric importance samplers tend to be far from the estimand $\cE$. 

The inspection of the $g$-specific ESS suggests that importance weights in the tail of nonparametric importance sampling density $\hat{q}^*_h(X_i)$ can be unstable. Especially because the kernel regression can poorly estimate the tail part, tail-controlling techniques \citep{owen2000safe,li2013two} can be helpful to address the issue.
}

We attribute the erratic $n\textrm{MSE}$ of nonparametric importance sampler in Figure~\ref{fig:exp-exp-results}(a) to the severe violation of the assumption, made as a basis of our theoretical analysis in Section~\ref{sec::thm}, that the natural configuration density $p$ has a compact support $\K\subset \R^d$. In this example, the tail of $p(x) \sim \exp(-x)$ decays even more slowly than the tail of $p(x) \sim \exp(-x^2)$ in the normal-normal example. The simulation experiment results for $d=2$ and 4 are not presented here, as they repeat the same pattern as for $d=1$. We note that the natural configuration density $p$ in the case study in Section~\ref{sec::emp::case_study} has the compact support, indicating that the assumption is still practical.

\subsection{Case Study}\label{sec::emp::case_study}
Wind energy is one of the fastest growing renewable energy sources \citep{you2017wind}. 
Yet, harvesting wind energy remains expensive, compared with fossil energy sources such as oil, coal, and natural gas, due to the high capital cost in installing wind turbines. A utility-scale wind turbine whose blade diameter is commonly greater than 100 ft typically costs more than one million U.S. dollars. Therefore, wind energy industry pays the utmost attention on ensuring the structural reliability of the wind turbine to prevent its failure \citep[e.g.,][]{moriarty2008,graf2017advances}. 

At the design stage of wind turbine, evaluating its reliability based on physical experiments is very limited due to the associated costs. Alternatively, the international standard, IEC  61400-1 \citep{IEC2005}, requires wind turbine manufacturers to use stochastic simulation models. For this purpose, the most widely used simulation models in the U.S. include TurbSim \citep{Jonkman2009} and FAST \citep{Jonkman2005}, which are developed and maintained by the National Renewable Energy Laboratory (NREL) of the U.S. Department of Energy. TurbSim simulates a 3-dimensional time marching wind profile, which becomes an input to FAST that, in turn, simulates a wind turbine's structural response to the wind. This case study focuses on two types of bending moments at the root of a turbine blade, namely, edgewise and flapwise bending moments, which represent two perpendicular structural responses of the blade root due to an external force or moment causing it to bend. We use the same benchmark turbine model \citep{jonkman2009_5MW} and simulator setup as \citet{moriarty2008} (see the references for the rest of the settings not provided below). 

In this case study, the input configuration $X$ is a 10-min average wind speed (unit: meter per second, m/s), which is fed into TurbSim. $X$ is sampled from the truncated Rayleigh distribution with the support of $[3,25]$ and the scale parameter $10\sqrt{2/\pi}$. The simulation output of interest, $V$, is the 10-min maximum bending moment (unit: killonewton meter, kNm) at the blade root, which is produced by FAST based on the simulated wind from TurbSim. Because $V$ is random even for a fixed $X$ due to the randomness of wind profile generated in TurbSim, we regard TurbSim and FAST together as a black-box stochastic simulation model.

To compare the nonparametric importance sampler proposed in this paper with a parametric importance sampler, we take a parametric method in \citet{choe2015importance} as a benchmark, which also approximates the optimal sampling density $q^*(x) \propto \sqrt{r(x)} \cdot p(x)$ in Lemma~\ref{lem::var}. We use the same simulation experiment setup therein: For the edgewise bending moment, we use $n=3600$, $m=600$, and $\xi = 9300$ kNm; for the flapwise bending moment, we use $n=9600$, $m=600$, and $\xi = 14300$ kNm. 

To model $r(x)$ using a pilot sample, $X$ is sampled $m=600$ times from a uniform distribution with the support of $[3, 25]$, and the corresponding $V$'s are generated from the NREL simulators. To build the parametric model of $r(x)$, the generalized additive model for location, scale and shape (GAMLSS) \citep{rigby2005} is fitted to the pilot sample. Specifically, the GAMLSS model assumes that the conditional distribution of $V$ given $X$ is a generalized extreme value distribution whose location and scale parameters are cubic spline functions of $X$ while the shape parameter is constant over $X$. The model parameters are estimated using the backfitting algorithm \citep{rigby2005}. 
To build the nonparametric model of $r(x)$, we fit the kernel regression model to the pilot sample with the Gaussian kernel and choose the smoothing bandwidth by cross-validation.

For both parametric and nonparametric importance samplers, we repeat estimating the failure probability $\cE =  P(V >\xi )$ 50 times (in contrast to 10,000 times in the numerical study in Section~\ref{sec::emp::num_study}). Recall that running the NREL simulators once takes about 1-min wall-clock time, implying that obtaining the pilot sample of size $m=600$ takes roughly 10 hours. We use the same pilot sample in all 50 replications, 
because repeating 50 times of the simulation experiment with $n-m = 3000$ or $9000$ itself requires several days of computation even if we use high-performance computing described in Section~\ref{sec::emp::num_study}. 

The parametric importance sampler in \citet{choe2015importance} uses the failure probability estimator 
\begin{equation}
\tilde{\cE}_{\hat{\theta}_m} = \frac{1}{n-m} \sum_{i=m+1}^n g(V_i)\frac{p(X_i)}{q^*_{\hat{\theta}_m}(X_i)},  \label{eq:param_IS_estimator_wo_pilot}
\end{equation}
 where the pilot sample is not used, compared with the estimator in \eqref{eq:param_IS_estimator}. In their procedure, the pilot sample is only used to build the model of $r(x)$. For fair comparison, we report both estimation results with and without using the pilot sample in the estimator. Note that nonparametric importance samplers for stochastic simulation models are not reported in the literature.

Table~\ref{tab:edge} summarizes the simulation experiment results for edgewise bending moments. We see that the standard errors of parametric importance samplers and those of nonparametric importance samplers are  not significantly different. 
Computational savings of both methods against CMC are remarkable and, at the same time, comparable with each other. 

\begin{table}[!hbt]
	\centering
	\caption{Estimation of the failure probability $\cE = P(V >\xi )$ for the edgewise bending moment $V$ and the threshold $\xi = 9300$ kNm}
	\label{tab:edge}
	\resizebox{\textwidth}{!}{\begin{tabular}{lccclccc}
			\hline
			& \multicolumn{3}{c}{Without the pilot sample} &  & \multicolumn{3}{c}{With the pilot sample} \\ \cline{2-4} \cline{6-8} 
			& Sample    & Standard error        & Comp.    &  & Sample   & Standard error       & Comp.   \\
			Method        & mean      & (95\% bootstrap CI)   & saving   &  & mean     & (95\% bootstrap CI)  & saving  \\ \hline
			Parametric    & 0.01005   & 0.00044               & 93\%     &  & 0.01016  & 0.00036              & 95\%    \\
			&           & (0.00036, 0.00051)    &          &  &          & (0.00030, 0.00042)   &         \\
			Nonparametric & 0.00998   & 0.00046               & 92\%     &  & 0.01010  & 0.00038              & 95\%    \\
			&           & (0.00034, 0.00056)    &          &  &          & (0.00029, 0.00047)   &         \\ \hline
	\end{tabular}}
	\\
	\raggedright
	{\footnotesize\textit{Note:} The computational (comp.) saving is $(n_{(CMC)} - n ) / n_{(CMC)}$, where $n_{(CMC)}= \hat{P}(1-\hat{P})/(\textrm{Standard error})^2$ is the theoretical sample size required for CMC to attain the same standard error when the true failure probability is equal to $\hat{P}$, which is the sample mean of the parametric importance samplers using the pilot sample. The 95\% bootstrap confidence interval (CI) is constructed based on 100,000 bootstrap replicates.
	} 
\end{table}

Table~\ref{tab:flap} shows the estimation results for flapwise bending moments, which convey the similar message with the results in Table~\ref{tab:edge}. Note that $\xi$ is set to roughly yield the similar failure probability $\cE = P(V >\xi )$ of 0.01 for both structural load types, because the magnitude of $\cE$ tends to impact the computational saving. 
Yet, we see that the computational saving of importance sampling over CMC for flapwise bending moments is, albeit substantial, not as large as that for edgewise bending moment. 
This is because the natural configuration density $p$ is not very different from the optimal sampling density $q^*$ for flapwise bending moments so that the benefit of changing the sampling density is not enormous.

\begin{table}[!hbt]
	\centering
	\caption{Estimation of the failure probability $\cE = P(V >\xi )$ for the flapwise bending moment $V$ and the threshold $\xi = 14300$ kNm}
	\label{tab:flap}
	\resizebox{\textwidth}{!}{\begin{tabular}{lccclccc}
		\hline
		& \multicolumn{3}{c}{Without the pilot sample} &  & \multicolumn{3}{c}{With the pilot sample} \\ \cline{2-4} \cline{6-8} 
		& Sample    & Standard error        & Comp.    &  & Sample   & Standard error       & Comp.   \\
		Method        & mean      & (95\% bootstrap CI)   & saving   &  & mean     & (95\% bootstrap CI)  & saving  \\ \hline
		Parametric    & 0.01037   & 0.00063               & 64\%     &  & 0.01079  & 0.00059              & 69\%    \\
		&           & (0.00046, 0.00078)    &          &  &          & (0.00043, 0.00073)   &         \\
		Nonparametric & 0.01061   & 0.00075               & 49\%     &  & 0.01101  & 0.00070              & 56\%    \\
		&           & (0.00057, 0.00090)    &          &  &          & (0.00053, 0.00084)   &         \\ \hline
	\end{tabular}}
\\
\raggedright
{\footnotesize\textit{Note:} Refer to the note of Table~\ref{tab:edge}.
}
\end{table}

This case study considers a relatively moderate failure probability $\cE = P(V >\xi )$ around 0.01, so we can use the given computational resource to have 50 replications that allow us to construct bootstrap confidence intervals on standard errors and compare the performances of methods. On the other hand, simulation experiments at the scale of a national lab may consider a more extreme event with a smaller probability. For such case, combining the 50 replications can lead to an estimate of a much smaller probability (in the order of 1e-05 or smaller as in \cite{choe2016}). Still, one can construct a confidence interval on the probability if only using the sample from the second stage \citep{Choe2017Uncertainty}.

\section{Discussion}\label{sec::disc}

We consider the problem of estimating the average output from a stochastic simulation model
and propose two-stage estimators using either a parametric approach or a nonparametric approach.
Theoretically, both estimators satisfy the oracle inequalities but they achieve the oracle variance asymptotically under different rates and assumptions.
As expected, the parametric approach needs a strong assumption but its variance converges to the oracle variance faster than the nonparametric approach. 
The nonparametric approach, however, requires weak assumptions but the variance reduction rate is not as fast as the parametric approach. 
Empirically, our numerical study confirmed the theoretical results, and our case study indicated that the proposed importance samplers perform well in practice, saving 50\%--95\% computational resources over a standard Monte Carlo estimator.

We note that
\cite{choe2015importance} investigated a parametric importance sampler for a stochastic simulation model, which is a special case (with $g(v) = 1(v\geq \xi)$) of our parametric importance sampler.
They use implicitly a two-stage method with the pilot sample being given \textit{a priori} for modeling $r(x)$ and do not use the pilot sample in the estimator. 
They neither consider nonparametric importance samplers nor perform a theoretical analysis on the convergence toward the optimal estimator. 
Our work focuses more on the formalization of the general two-stage strategy and the theoretical foundation
of importance sampling for stochastic simulation models.
Because we obtain a concrete convergence rate of the estimator, 
we enable the optimal choice of the pilot sample size. 


In what follows we discuss possible future research directions. 
\begin{itemize}
\item {\bf Manifold support case.} 
In general, when the dimension of the configuration $d$ is large, nonparametric importance sampler in Section~\ref{sec::NIS} will not work due to the curse of dimensionality (the slow convergence rate). 
However, if the support of $q^*$ is concentrated around a lower dimensional manifold, the nonparametric sampler may still work because a fast convergence rate
of a nonparametric estimator is possible \citep{balakrishnan2013cluster,chen2016generalized}. So we may be able to design
a modified nonparametric importance sampling procedure that achieves oracle variance
much faster than the rate in Corollary~\ref{cor::mn}.
The construction of such a procedure is left as a future work.

\item {\bf Multi-stage sampling.}
In this paper we only consider splitting the computational budget into two stages.
We can generalize this idea into a $k$-stage sampling procedure,
where at each stage, we use samples in all previous stages to
design our estimator and sampler for the current stage \citep[e.g.,][]{choe2017information}.
In this case, the allocation problem becomes more complicated since 
we may assign different sample sizes to different stages.  
Also, the number of stage $k$ will be another quantity that we want to optimize. 
Because the two-stage approach is a special case of a multi-stage sampling procedure, 
the latter will have a higher variance reduction rate than the proposed methods in this paper.

\item {\bf Confidence interval.}
In the current paper, we focus on the construction of an estimator of $\cE$.
In practice, we often report not only a point estimator but also
a confidence interval attached to it.
Here we briefly describe two potential methods of constructing the confidence interval. 
The first method is to derive asymptotic normality of $\hat{\cE}$ and then find a consistent variance estimator.
Note that this is a non-trivial task because when we use a two-stage approach,
the observations are no longer IID. 
Moreover, estimating the variance could be another challenging task. 
The other approach is to use the bootstrap \citep{efron1982jackknife,efron1992bootstrap} to obtain a confidence interval. 
If we choose to use the bootstrap, we need to prove the validity of such a bootstrap procedure.

{\color{black}\item {\bf Multiple estimands of interest.} This paper considers the single quantity to be estimated, $\cE = \mathbb{E}(g(V(X)))$, in \eqref{eq::target} based on the output $V(X)$ from a stochastic simulation model. Extending the current work to optimally estimating multiple quantities of interest warrants further investigation. Such extension for stochastic simulation models can build upon this line of research around deterministic simulation models. For example, we can consider importance sampling of the union of target events \citep{owen2019importance}. Also,  we can recycle existing simulation results (e.g., from the pilot stage or both stages of estimating another quantity) by recomputing importance weights \citep{cornuet2012adaptive}. These approaches can benefit from the classical idea of safe and effective importance sampling \citep{owen2000safe}.}

\end{itemize}



{\color{black}\section*{Acknowledgements}
This work was partially supported by the NSF grant CMMI-1824681 and DMS 1810960 and NIH grant U01-AG016976.}

\bibliographystyle{abbrvnat}
\bibliography{NPIS} 

\appendix

\section{Proofs}
\begin{proof}[ of Lemma~\ref{lem::var}]

Now by the following variance formula:
$$
{\sf Var} (Y) = \mathbb{E}({\sf Var}(Y|X)) + {\sf Var} (\mathbb{E}(Y|X))
$$
and choose $Y = g(V_i)\frac{p(X_i)}{q(X_i)}$ and $X = X_i$, we have 
\begin{equation}
\begin{aligned}
{\sf Var} \left(g(V_i)\frac{p(X_i)}{q(X_i)}\right) 
& = \mathbb{E}\left({\sf Var}\left(g(V_i)|X_i\right)\frac{p^2(X_i)}{q^2(X_i)}\right) + 
{\sf Var} \left(r^\dagger(X_i) \frac{p(X_i)}{q(X_i)}\right)\\
& = \mathbb{E}\left(\left(r(X_i)-r^{\dagger 2}(X_i)\right)\frac{p^2(X_i)}{q^2(X_i)}\right) \\
&\qquad\qquad+ \mathbb{E}\left(r^{\dagger2}(X_i) \frac{p^2(X_i)}{q^2(X_i)}\right) - \mathbb{E}^2\left(r^\dagger(X_i) \frac{p(X_i)}{q(X_i)}\right)\\
& = \mathbb{E}\left(r(X_i) \frac{p^2(X_i)}{q^2(X_i)}\right) - \mathbb{E}^2(r^\dagger(X^*)),
\end{aligned}
\label{eq::var1}
\end{equation}
when $q(x)=0$ implies $r^\dagger(x)p(x)=0$.
Note that  $X^*$ is from density $p$. 
Thus, the sampling density $q$ affects the variance only via the quantity $\mathbb{E}\left(r(X_i) \frac{p^2(X_i)}{q^2(X_i)}\right)$.

The quantity $\mathbb{E}\left(r(X_i) \frac{p^2(X_i)}{q^2(X_i)}\right)$ has a lower bound 
from the Cauchy-Schwarz inequality:
\begin{equation}
\begin{aligned}
\mathbb{E}\left(r(X_i) \frac{p^2(X_i)}{q^2(X_i)}\right) & = \int r(x) \frac{p^2(x)}{q(x)} dx\\
& = \int \left(\sqrt{r(x)}\frac{p(x)}{\sqrt{q(x)}}\right)^2 dx \cdot \underbrace{\int (\sqrt{q(x)})^2 dx}_{=1} \\
&\geq \left(\int \sqrt{r(x)} p(x)dx\right)^2 \\
& = \mathbb{E}^2\left(\sqrt{r(X^*)}\right).
\end{aligned}
\label{eq::var2}
\end{equation}
And the equality holds when $\sqrt{r(x)}\frac{p(x)}{\sqrt{q(x)}} \propto \sqrt{q(x)}$,
which implies the optimal sampling density is
$$
q^*(x) \propto \sqrt{r(x)} \cdot p(x).
$$

Thus, when we choose the sampling density to be $q^*$, by equation \eqref{eq::var1} and \eqref{eq::var2},
the variance 
\begin{align*}
{\sf Var} \left(\hat{\cE}_{q^*}\right) &= \frac{1}{n}\mathbb{E}\left(r(X_i) \frac{p^2(X_i)}{q^2(X_i)}\right) - \mathbb{E}^2(r^\dagger(X^*))\\
& = \frac{1}{n}\left(\mathbb{E}^2\left(\sqrt{r(X^*)}\right)-\mathbb{E}^2\left(r^\dagger(X^*)\right)\right).
\end{align*}

\end{proof}

\begin{proof}[ of Theorem~\ref{thm::Prate}]
By assumptions (P1--2) and the M-estimation theory \citep{van1996weak,van2000asymptotic}, 
$$
\|\hat{\theta}_m - \theta_0\| = O_P\left(\sqrt{\frac{1}{m}}\right),
$$
where $\theta_0$ is the parameter such that $r(x)=r_{\theta_0}(x)$.

Now by assumption (P3),
\begin{align*}
\sup_{x\in\K}\|r_{\hat{\theta}_m}(x)-r(x)\| &= \sup_{x\in\K}\|r_{\hat{\theta}_m}(x)-r_{\theta_0}(x)\|\\
& \leq \sup_{x\in\K} \left\|(\hat{\theta}_m-\theta_0)\cdot L_0 \right\|\\
& = O_P\left(\sqrt{\frac{1}{m}}\right),
\end{align*} 
which proves the result.
\end{proof}

\begin{proof}[ of Theorem~\ref{thm::P_var}]

For our estimator $\hat{\cE}_{\hat{\theta}_m}$, we decompose it into two parts
$$
\hat{\cE}_{\hat{\theta}_m} = A_m + B_n,
$$
where 
\begin{align*}
A_m & = \frac{1}{n}\sum_{i=1}^m g(V_i)\frac{p(X_i)}{q_0(X_i)},\\
B_n&= \frac{1}{n}\sum_{i=m+1}^n g(V_i)\frac{p(X_i)}{q^*_{\hat{\theta}_m}(X_i)},
\end{align*}
Thus, 
\begin{equation}
{\sf Var} \left(\hat{\cE}_{\hat{\theta}_m}\right) = {\sf Var} (A_m + B_n) = {\sf Var} (A_m) + {\sf Var} (B_n) + 2{\sf Cov}(A_m,B_n).
\label{eq::cov1}
\end{equation}
Note that 
\begin{equation}
\begin{aligned}
\mathbb{E}(A_m )&= \frac{m}{n}\mathbb{E}_{X_i\sim q_0} \left(g(V_i)\frac{p(X_i)}{q_0(X_i)}\right) = \frac{m}{n}\cE\\
\mathbb{E}(B_n )&= \frac{n-m}{n}\cdot\mathbb{E}_{X_i\sim q^*_{\hat{\theta}_m}} \left(g(V_i)\frac{p(X_i)}{q^*_{\hat{\theta}_m}(X_i)}\right) \\
&= \frac{n-m}{n}\cdot\mathbb{E}_{X_i\sim q^*_{\hat{\theta}_m}}\left(\mathbb{E} \left(g(V_i)\frac{p(X_i)}{q^*_{\hat{\theta}_m}(X_i)}|q^*_{\hat{\theta}_m}\right)\right) \\
&= \frac{n-m}{n}\cdot \cE.
\end{aligned}
\label{eq::cov2}
\end{equation} 

We first bound the covariance. 
Let $\mathcal{D}_m = \{(X_1,V_1),\cdots,(X_m,V_m)\}$ be the collection of the first part of the data. 
Then
\begin{equation}
\begin{aligned}
{\sf Cov}(A_m,B_n) & = \mathbb{E}(A_m B_n) - \mathbb{E}(A_m)\mathbb{E}(B_n)\\
&= \mathbb{E}(A_m \mathbb{E}(B_n|\mathcal{D}_m)) - \frac{(n-m)\cdot m}{n^2}\cdot \cE^2\\
&= \mathbb{E}(A_m \mathbb{E}(B_n|q^*_{\hat{\theta}_m})) - \frac{(n-m)\cdot m}{n^2}\cdot \cE^2\\
&= \mathbb{E}\left(A_m \cdot \frac{n-m}{n}\cdot \cE\right) - \frac{(n-m)\cdot m}{n^2}\cdot \cE^2\\
& = \frac{(n-m)\cdot m}{n^2}\cdot \cE^2 - \frac{(n-m)\cdot m}{n^2}\cdot \cE^2\\ 
&= 0 .
\end{aligned}
\label{eq::cov3}
\end{equation}
Therefore, we only need to focus on the variance of each part.

Let $V_{\min} = \left(\mathbb{E}^2\left(\sqrt{r(X^*)}\right) - \mathbb{E}^2(r^\dagger(X^*))\right)$ be the minimal
variance under the optimal sampling density.
By Lemma~\ref{lem::var}, 
\begin{equation}
\begin{aligned}
 {\sf Var} (A_m)  &= \frac{m}{n^2} \left(\mathbb{E}\left(r(X_i) \frac{p^2(X_i)}{q_0^2(X_i)}\right) - \mathbb{E}^2(r^\dagger(X^*))\right)\\
 & = \frac{m}{n^2} V_{\min} + \frac{m}{n^2}\left(\mathbb{E}\left(r(X_i) \frac{p^2(X_i)}{q_0^2(X_i)}\right)-\mathbb{E}^2\left(\sqrt{r(X^*)}\right)\right)\\
 & = \frac{m}{n^2} V_{\min} + \frac{m}{n^2} V_{q_0}.
 \end{aligned}
 \label{eq::cov4}
\end{equation}

And the variance of the second part is
\begin{equation}
\begin{aligned}
{\sf Var} (B_n) &=  \mathbb{E}\left({\sf Var} (B_n|\mathcal{D}_m)\right) + {\sf Var}\left(\mathbb{E}(B_n|\mathcal{D}_m)\right)\\
&=\mathbb{E}\left({\sf Var} (B_n|\mathcal{D}_m)\right) + \underbrace{{\sf Var}\left(\cE\right)}_{=0}\\
&= \frac{n-m}{n^2}\cdot \mathbb{E}\left(\mathbb{E}\left(r(X_i) \frac{p^2(X_i)}{{q}^{*2}_{\hat{\theta}_m}(X_i)}|{q}^*_{\hat{\theta}_m}\right) - \mathbb{E}^{2}(r^\dagger(X^*))\right)\\
& = \frac{n-m}{n^2}\cdot \mathbb{E}\left(\mathbb{E}\left(r(X^*) \frac{p(X^*)}{{q}^*_{\hat{\theta}_m}(X^*)}|{q}^*_{\hat{\theta}_m}\right)\right) - \frac{n-m}{n^2}\cdot  \mathbb{E}^2(r^\dagger(X^*)).
\end{aligned}
\label{eq::cov5}
\end{equation} 
So the key part is in the quantity $\mathbb{E}\left(\mathbb{E}\left(r(X^*) \frac{p(X^*)}{{q}^*_{\hat{\theta}_m}(X^*)}|{q}^*_{\hat{\theta}_m}\right)\right)$.
By Theorem~\ref{thm::Prate} we have
$$
\sup_{x\in\K}\|r_{\hat{\theta}_m}(x)  - r(x)\| = O_P\left(\sqrt{\frac{1}{m}}\right),
$$
which implies 
$$
q^*_{\hat{\theta}_m}(x) - q^*(x) = \Delta_m\cdot p(x),
$$ 
where $\Delta_m = O_P\left(\sqrt{\frac{1}{m}}\right)$.
Thus,
\begin{equation}
\begin{aligned}
\mathbb{E}\left(r(X^*) \frac{p(X^*)}{{q}^*_{\hat{\theta}_m}(X^*)}|{q}^*_{\hat{\theta}_m}\right) & = \int r(x) \frac{p^2(x)}{{q}^*_{\hat{\theta}_m}(x)}dx\\
& = \int r(x) \frac{p^2(x)}{q^*(x) + \Delta_m\cdot p(x)}dx\\
& = \int r(x) \frac{p^2(x)}{q^*(x)} dx + O\left(\Delta_m\right)\\
& = \mathbb{E}^2\left(\sqrt{r(X^*)}\right) + O_P\left(\sqrt{\frac{1}{m}}\right).
\end{aligned}
\end{equation} 
{\color{black}Note that in the above equation,
the constant in the $O(\Delta_m)$
term is
$$
\int r(x) p^3(x)/q^*(x)^2dx = \int r(x) p(x)/[C^2\cdot r(x)]dx = \frac{1}{C^2}\int p(x) dx = \frac{1}{C^2},
$$
where $C$ is the constant from equating $q^*(x) = C \sqrt{r(x)} p(x)$. 
The assumption (P1) implies that 
the ratio $\frac{p(x)}{q^*(x)} = \frac{1}{C\cdot r^{-1/2}(x)}$
is uniformly bounded.
Thus, the higher order terms are bounded.}

Putting this back to equation \eqref{eq::cov5}, we obtain
\begin{align*}
{\sf Var} (B_n) &= \frac{n-m}{n^2} \left(\mathbb{E}^2\left(\sqrt{r(X^*)}\right) - \mathbb{E}^2(r^\dagger(X^*))\right) + \frac{n-m}{n^2}\E \left(O_P\left(\sqrt{\frac{1}{m}}\right)\right)\\
& = \frac{n-m}{n^2} V_{\min} +\frac{n-m}{n^2} O\left(\sqrt{\frac{1}{m}}\right).
\end{align*} 
Note that $\E \left(O_P\left(\sqrt{\frac{1}{m}}\right)\right)= O\left(\sqrt{\frac{1}{m}}\right)$
because the random quantity in the $O_P$ term is from $\Delta_m$
and is from the difference $|\hat{\theta}_m - \theta_0|$, which is the absolute value of an asymptotic normal distribution 
so the expectation of the $O_P$ leads to the same convergence rate.



Now putting altogether, we obtain
\begin{equation}
\begin{aligned}
{\sf Var} \left(\hat{\cE}_{\hat{\theta}_m}\right) &= \frac{m}{n^2} V_{\min} + \frac{m}{n^2} V_{q_0} + \frac{n-m}{n^2} V_{\min} +\frac{n-m}{n^2} O\left(\sqrt{\frac{1}{m}}\right)\\
& = \frac{1}{n}V_{\min} + \frac{1}{n^2}\left(m\cdot V_{q_0} + (n-m)\cdot O\left(\sqrt{\frac{1}{m}}\right)\right).
\end{aligned}
\label{eq::t_opt1}
\end{equation}

\end{proof}

\begin{proof}[ of Theorem~\ref{thm::rate}]
Recall that $\hat{r}_h(x) = \frac{\sum_{i=1}^m Y_i K\left(\frac{x-X_i}{h}\right)}{\sum_{i=1}^m K\left(\frac{x-X_i}{h}\right)}$
and $r(x) = \mathbb{E}(Y_i|X_i=x)$.
Define the following two quantities 
\begin{equation}
\begin{aligned}
\tilde{r}_h(x) &= \frac{\frac{1}{mh^d}\sum_{i=1}^m Y_i K\left(\frac{x-X_i}{h}\right)}{q_0(x)},\\
\overline{r}_h(x) & = \frac{\E \left(\frac{1}{h^d}Y_i K\left(\frac{x-X_i}{h}\right)\right)}{q_0(x)}.
\end{aligned}
\end{equation}
We can bound the difference $\sup_{x\in\K}\|\hat{r}_h(x)-r(x)\|$ by
\begin{equation}
\begin{aligned}
\sup_{x\in\K}\|\hat{r}_h(x)&-r(x)\| = \sup_{x\in\K}\|\hat{r}_h(x)-\tilde{r}_h(x)+\tilde{r}_h(x)-\overline{r}_h(x)+\overline{r}_h(x)-r(x)\|\\
&\leq \sup_{x\in\K}\|\hat{r}_h(x)-\tilde{r}_h(x)\|+\sup_{x\in\K}\|\tilde{r}_h(x)-\overline{r}_h(x)\|+\sup_{x\in\K}\|\overline{r}_h(x)-r(x)\|.
\end{aligned}
\label{eq::rate::1}
\end{equation} 
Now we separately bound each term.

The first term $\sup_{x\in\K}\|\hat{r}_h(x)-\tilde{r}_h(x)\|$ involves 
the difference between $q_0(x)$ and $\frac{1}{mh^d}\sum_{i=1}^m K\left(\frac{x-X_i}{h}\right)= \hat{q}_m(x)$,
which is the difference between the kernel density estimator (KDE) $\hat{q}_m(x)$ and $q_0(x)$. 
By assumption (N2) and (K1--2), it is known in the literature that 
$$
\sup_{x\in\K}\|\hat{q}_m(x)-q_0(x)\| = O(h^2)+O_P\left(\sqrt{\frac{\log m}{mh^d}}\right);
$$
see, e.g., Lemma 5 in \cite{chen2015optimal} and Lemma 10 in \cite{chen2015asymptotic}.
Thus, 
\begin{equation}
\sup_{x\in\K}\|\hat{r}_h(x)-\tilde{r}_h(x)\| = O(h^2)+O_P\left(\sqrt{\frac{\log m}{mh^d}}\right).
\label{eq::rate::2}
\end{equation}

For the second term, it equals to 
\begin{align*}
\sup_{x\in\K}\|\tilde{r}_h(x)-\overline{r}_h(x)\| 
&= \sup_{x\in\K}\left\|\frac{\frac{1}{mh^d}\sum_{i=1}^m Y_i K\left(\frac{x-X_i}{h}\right)}{q_0(x)}-
\frac{\E \left(\frac{1}{h^d}Y_i K\left(\frac{x-X_i}{h}\right)\right)}{q_0(x)}\right\|\\
&=\sup_{x\in\K}\left\| \frac{1}{q_0(x)}\left(\frac{1}{mh^d}\sum_{i=1}^m Y_i K\left(\frac{x-X_i}{h}\right)-\E \left(\frac{1}{h^d}Y_i K\left(\frac{x-X_i}{h}\right)\right)\right) \right\|.
\end{align*}
Now using Theorem 2.3 in \cite{Gine2002} and assumption (N1) and (K1--2), 
we can bound
$$
\sup_{x\in\K}\left\|\frac{1}{mh^d}\sum_{i=1}^m Y_i K\left(\frac{x-X_i}{h}\right)-\E \left(\frac{1}{h^d}Y_i K\left(\frac{x-X_i}{h}\right)\right)\right\| = O_P\left(\sqrt{\frac{\log m}{mh^d}}\right).
$$
Assumption (N2) implies that the density $q_0(x)$ is lower bounded by $q_{\min}$.
Thus, we obtain the bound 
\begin{equation}
\begin{aligned}
\sup_{x\in\K}\|\tilde{r}_h(x)-\overline{r}_h(x)\|  &\leq \frac{1}{q_{\min}}\sup_{x\in\K}\left\|\frac{1}{mh^d}\sum_{i=1}^m Y_i K\left(\frac{x-X_i}{h}\right)-\E \left(\frac{1}{h^d}Y_i K\left(\frac{x-X_i}{h}\right)\right)\right\| \\
&=  O_P\left(\sqrt{\frac{\log m}{mh^d}}\right).
\end{aligned}
\label{eq::rate::3}
\end{equation}

The third term $\sup_{x\in\K}\|\overline{r}_h(x)-r(x)\|$ involves the bias in nonparametric regression
which is known to be at rate $O(h^2)$ under assumption (N2).
Based on this rate and equations \eqref{eq::rate::2} and \eqref{eq::rate::3}, by equation \eqref{eq::rate::1}
we obtain
$$
\sup_{x\in\K}\|\hat{r}_h(x)-r(x)\|= O(h^2)+O_P\left(\sqrt{\frac{\log m}{mh^d}}\right),
$$
which is the desired result.

\end{proof}

\begin{proof}[ of Theorem~\ref{thm::NP_var}]
This proof follows the same way as the proof of Theorem~\ref{thm::P_var}:
we decompose $\hat{\cE}_{h^*} = A_m + B_n$
and control the variance of $A_m$ and $B_n$ and show that the covariance is $0$. 

The only difference is in the variance of $B_n$. 
Because the estimation error now becomes (see equation \eqref{eq::unif1})
$$
\sup_{x\in\K}\|\hat{r}_{h^*}(x)-r(x)\|= O_P\left(\left(\frac{\log m}{m}\right)^{\frac{2}{d+4}}\right),
$$
the variance
\begin{align*}
{\sf Var} (B_n) &= \frac{n-m}{n^2} \left(\mathbb{E}^2\left(\sqrt{r(X^*)}\right) - \mathbb{E}^2(r^\dagger(X^*))\right) + \frac{n-m}{n^2} O\left(\left(\frac{\log m}{m}\right)^{\frac{2}{d+4}}\right)\\
& = \frac{n-m}{n^2} V_{\min} +\frac{n-m}{n^2} O\left(\left(\frac{\log m}{m}\right)^{\frac{2}{d+4}}\right).
\end{align*}
Thus, the total variance is
\begin{align*}
{\sf Var}\left(\hat{\cE}_{h^*}\right) & = {\sf Var} (A_m)+{\sf Var} (B_n)+ \underbrace{2{\sf Cov} (A_m,B_n)}_{=0}\\
&= \frac{m}{n^2} V_{\min} + \frac{m}{n^2} V_{q_0} + \frac{n-m}{n^2} V_{\min} +\frac{n-m}{n^2} O\left(\left(\frac{\log m}{m}\right)^{\frac{2}{d+4}}\right)\\
& = \frac{1}{n}V_{\min} + \frac{1}{n^2}\left(m\cdot V_{q_0} + (n-m)\cdot O\left(\left(\frac{\log m}{m}\right)^{\frac{2}{d+4}}\right)\right),
\end{align*}
which proves the desired result.
\end{proof}


\end{document}